\documentclass[aps,pre,twocolumn,superscriptaddress,amsmath,amssymb]{revtex4}

\usepackage{graphicx,dcolumn,txfonts}

\begin{document}

\title{Temporal Networks}

\author{Petter Holme}
\affiliation{IceLab, Department of Physics, Ume{\aa} University, 901 87 Ume{\aa}, Sweden}
\affiliation{Department of Energy Science, Sungkyunkwan University, Suwon 440--746, Korea}
\affiliation{Department of Sociology, Stockholm University, 106 91 Stockholm, Sweden}

\author{Jari Saram\"aki}
\affiliation{Department of Biomedical Engineering and Computational Science, School of Science, Aalto University, 00076 Aalto, Espoo, Finland}

\begin{abstract}
A great variety of systems in nature, society and technology---from the web of sexual contacts to the Internet, from the nervous system to power grids---can be modeled as graphs of vertices coupled by edges. The network structure, describing how the graph is wired, helps us understand, predict and optimize the behavior of dynamical systems. In many cases, however, the edges are not continuously active. As an example, in networks of communication via email, text messages, or phone calls, edges represent sequences of instantaneous or practically instantaneous contacts. In some cases, edges are active for non-negligible periods of time: e.g., the proximity patterns of inpatients at hospitals can be represented by a graph where an edge between two individuals is on throughout the time they are at the same ward. Like network topology, the temporal structure of edge activations can affect dynamics of systems interacting through the network, from disease contagion on the network of patients to information diffusion over an e-mail network. In this review, we present the emergent field of temporal networks, and discuss methods for analyzing topological and temporal structure and models for elucidating their relation to the behavior of dynamical systems. In the light of traditional network theory, one can see this framework as moving the information of \textit{when} things happen from the dynamical system on the network, to the network itself. Since fundamental properties, such as the transitivity of edges, do not necessarily hold in temporal networks, many of these methods need to be quite different from those for static networks. The study of temporal networks is very interdisciplinary in nature. Reflecting this, even the object of study has many names---temporal graphs, evolving graphs, time-varying graphs, time-aggregated graphs, time-stamped graphs, dynamic networks, dynamic graphs, dynamical graphs, and so on. This review covers different fields where temporal graphs are considered, but does not attempt to unify related terminology---rather, we want to make papers readable across disciplines.
\end{abstract}

\maketitle

\tableofcontents

\section{Introduction}

To get an overview of a large, integrated system, one needs to zoom out from the details. For many systems, from the Internet to the metabolism, from the proteome to the web of sexual contacts, an easy way of doing this is representing the system as a graph. A graph is a mathematical object consisting of a set of vertices, the units of the system, and a set of edges, the pairs of vertices that are interacting with each other. Usually, such networks are the infrastructure of some dynamical system---data-packet traffic on the Internet, disease spreading on social networks, etc.---and this dynamical system is what we are really interested in. The advantage of modeling the system as a graph is that we can say much about the behavior of the dynamical system without studying the actual dynamics at all. We can estimate how much one part of the network influences another; how well the network is optimized with respect to the dynamical system; which vertices play similar roles in the system's operation; and so on~\cite{newman_2010,barrat_etal_2008,jackson,easley_kleinberg}. Sometimes such a crude modeling framework can be made more powerful if one extends it to include additional levels of detail, for example edge weights in weighted networks~\cite{barrat_etal_2004}, or the position of vertices in spatial networks~\cite{barthelemy}. In this review, we consider an additional dimension---time---and discuss temporal networks, where the times when edges are active are an explicit element of the representation. Until recently, in most network studies, the time dimension has been projected out by aggregating the contacts between vertices to (sometimes weighted) edges, even in cases when detailed information on the temporal sequences of contacts or interactions would have been available. Sometimes the solution has been to segment the data into adjacent time windows where contacts are aggregated into edges, and then study the time evolution of the network structure in these windows. Such an approach does not cover all aspects of the temporal structure of contact patterns. For example, the edges between vertices of temporal networks need not be transitive. In static networks, whether directed or not, if A is directly connected to B and B is directly connected to C, then A is indirectly connected to C via a path over B. However, in temporal networks, if the edge (A,B) is active only at a later point in time than the edge (B,C), then A and C are disconnected, as nothing can propagate from A via B to C (Fig.~\ref{fig:reachability}). Thus, the time ordering can matter a lot, and as we shall see below, the timings of connections and their correlations do have effects that go beyond what can be captured by static networks. Accordingly, the main focus of this review is on methods that do not ignore the consequences of the time ordering by e.g.\ projecting out the interaction times.

When one studies a network, it is usually not the network itself (the vertices and edges) that is the object of study. Rather one wants to investigate a dynamical system \textit{on} the network. In traditional network modeling one separates the underlying static network and the dynamical system on the network. Compared to this picture, temporal network approaches moves information about \textit{when} things happen from the dynamical system to the network, the underlying structure on which the dynamics happen. Systems suitable to be modeled as temporal networks are everywhere. The flow of information via e-mail messages, mobile telephone calls, and social media is one such system that has recently attracted much attention. Likewise, detailed understanding of the spreading dynamics of some electronic and biological viruses calls for taking the properties of the underlying contact sequences into account. Studies of many networks in the life sciences---from activation sequences of genetic regulation to time-domain features of functional brain networks---may benefit from the temporal graph approach. Food webs and other networks of species evolve in time with environmental conditions that are to some extent a result of which species are present. This type of feedback fits the temporal-network framework. Another example is self-assembled networks of wireless devices and other distributed computing systems.

\begin{figure}
\includegraphics[width=\linewidth]{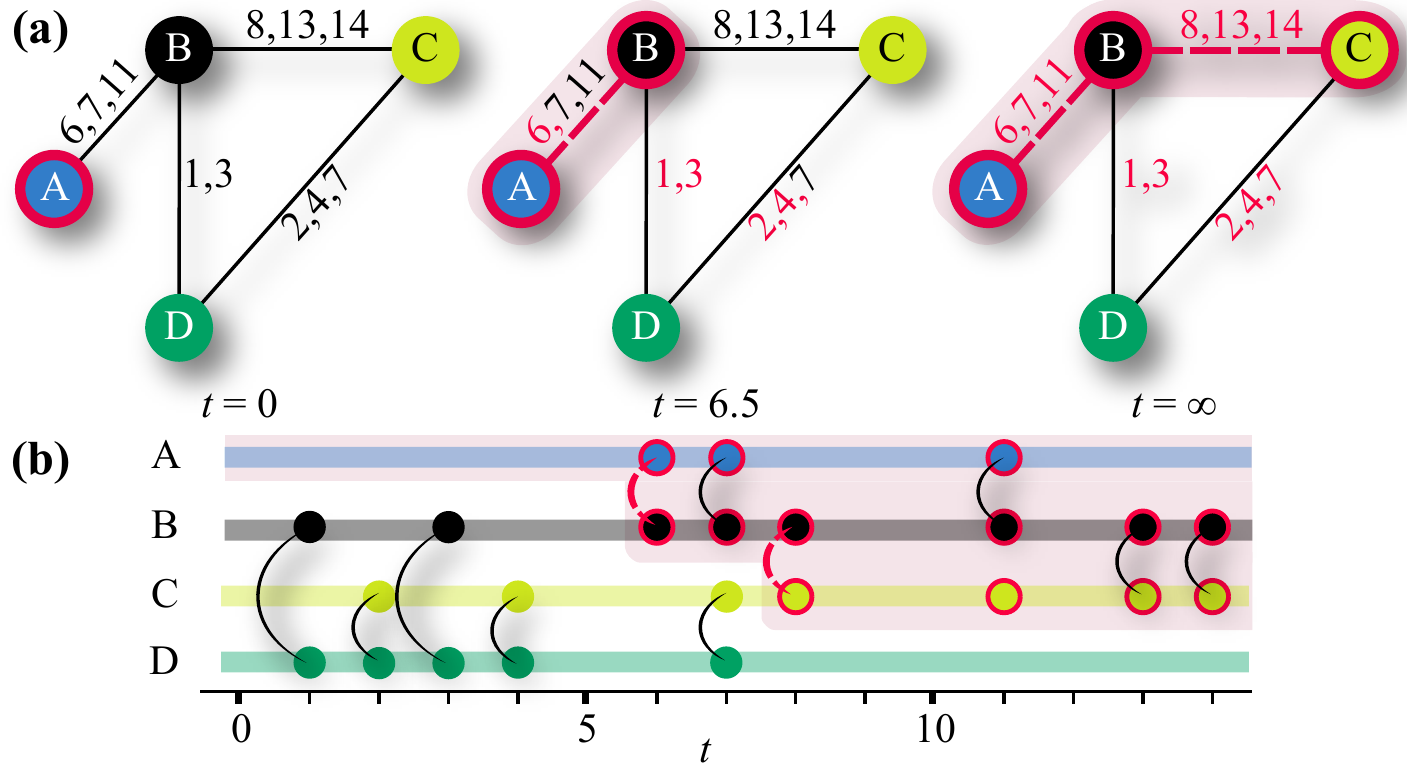}
\caption{Illustration of the reachability issue and the intransitivity of temporal networks (more specifically a contact sequence). In (a), the times of the contacts between vertices A--D are indicated on the edges. Assume that, for example, a disease starts spreading at vertex A and spreads further as soon as a contact occurs. The dashed lines and vertices show this spreading process for four different times. The spreading will not continue further than what is indicated in the $t=\infty$ picture, i.e.\ D cannot get infected. However, if the spreading started at vertex D, the entire set of vertices would eventually be infected. Aggregating the edges into one static graph cannot capture this effect that arises from the time ordering of contacts. Panel (b) visualizes the same situation by showing the temporal dimension explicitly. The colors of the lines in (b) matches the vertex colors in (a).}
\label{fig:reachability}
\end{figure}

\begin{figure*}
\includegraphics[width=0.7\linewidth]{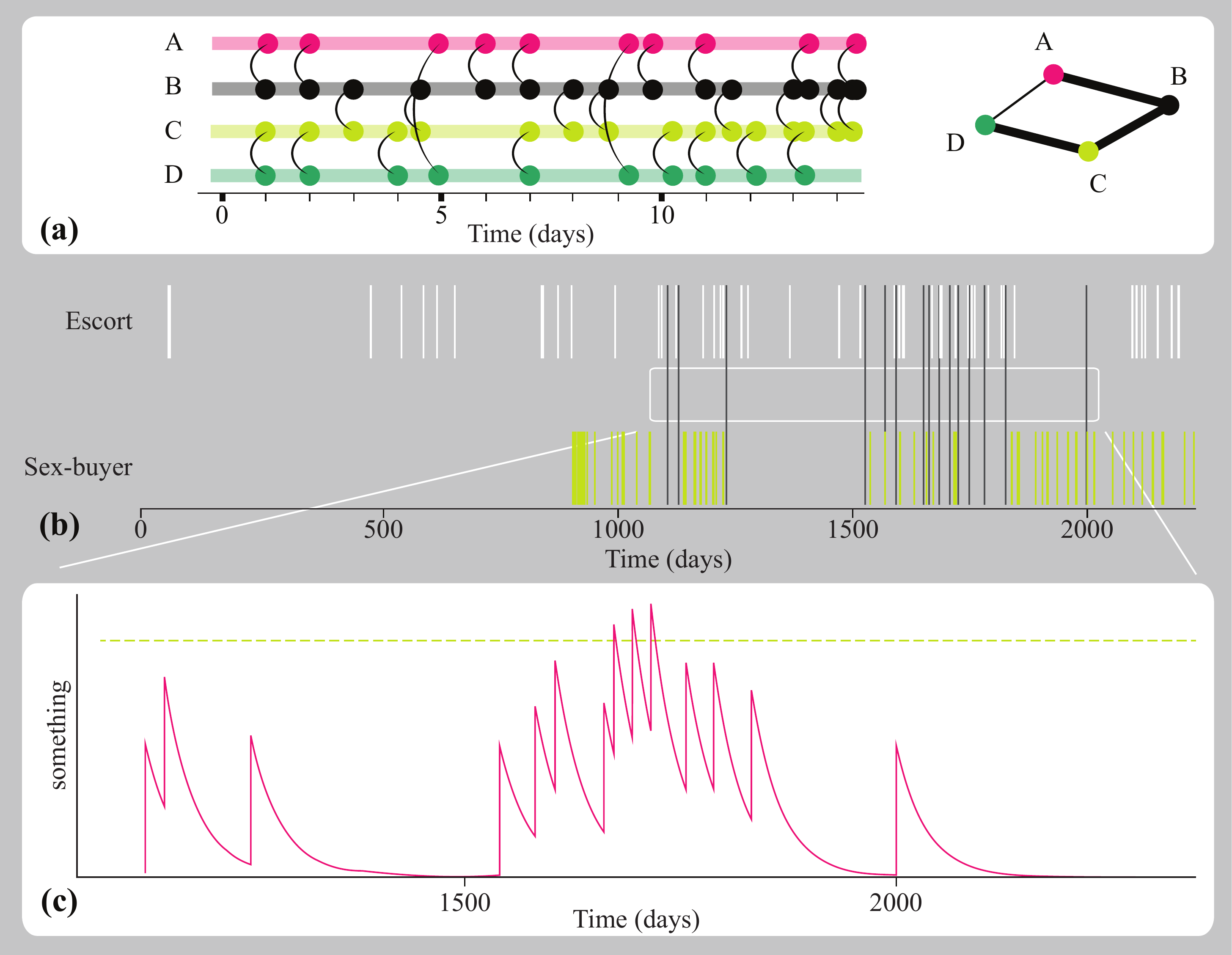}
\caption{The limits of applicability of aggregated contact sequences, in the context of spreading dynamics. Panel (a) shows a schematic contact sequence (similarly to Fig.~\ref{fig:reachability}(b)) that would be fairly well modeled as an aggregated weighted graph assuming a standard random contact process, as seen to the right. Panel (b) displays a real-world contact sequence involving two vertices in a real network of escorts and sex-sellers~\cite{rocha_etal_2010}. The vertical lines show the times when the individuals are active in the data, while a line connecting the two individuals indicates a contact between them. Both the behavior of the individuals and the activity of the edge between them are bursty, with periods of intense activity followed by silent periods. (c) shows a hypothetical dynamics where one of the individuals in (b) gets a dose of, for example, a pathogen from the other individual at every contact, and the concentration of the pathogen decays exponentially. 
If the individual becomes sick when the pathogen concentration reaches a threshold (the horizontal, dashed line), then bursty dynamics would bring the level over this threshold. On the contrary, for more regular contact dynamics such as those in panel (a), it would have time to decay below the threshold.}
\label{fig:limits_of_temporal_models}
\end{figure*}

In general, when is temporal networks a suitable framework for analysis and modeling? Just like for static complex networks, the system under study should consist of agents that interact pairwise, so that the interactions have both some degree of randomness and some regularity (i.e., there is some structure). We also need to require similar properties for the temporal structures---they should not be too random or too regular in order to fit the framework. On one hand, one will always lose information when projecting a temporal network structure to a static graph (see Fig.~\ref{fig:reachability} for an illustration). On the other hand, in some cases, this loss of information is probably too insignificant to make up for the more complicated analysis and modeling needed for the temporal graph approach. See Fig.~\ref{fig:limits_of_temporal_models}(a) for an illustration of a contact sequence that is fairly well modeled by a weighted graph with the assumption that contact times are random, with a frequency proportional to the edge weight. The dynamical system of interest on the network matters too---different systems can respond differently to a specific temporal structure. For a thought experiment, consider the empirical bursty contact pattern plotted in Fig.~\ref{fig:limits_of_temporal_models}(b). Assume that a contact triggers an increase of something (say, the concentration of a virus in the blood) in one of the vertices involved, which then decreases exponentially. Further, assume that the person gets sick and infectious if the virus concentration reaches a critical level. Then, bursty edge dynamics~\cite{kleinberg,kumar_etal,barabasi,eckmann_etal,zimo} could be of crucial importance for that something to propagate through the network. In a situation with more uncorrelated or evenly distributed times of contact, the virus concentration would have time to fall below the dangerous level between the contacts. Thus for such a dynamical system, bursty edge activity would play a far more important role than for a system where the dynamics can be modeled as a branching process~\cite{kimmel_axelrod}, as is the case for many network-based models of disease spreading.

A special case of the requirement that a system should have temporal structure for it to suit a temporal-network framework, relates to time scales~\cite{gautreau_etal}. If the dynamical system on the network is too rapid compared to the dynamics of the contacts, or when edges are active, then there is no need to model the system as a temporal network. One example is the Internet where the data packets travel much faster than the topology changes. In summary, if the system is temporally and topologically connected in a way that affects the dynamics of interest, then temporal networks may be an optimal theoretical framework.

The study of temporal networks is very much an interdisciplinary field, where much of the development has been taking place in parallel, seemingly without much communication between the different disciplines. This is reflected in a tremendous amount of overlapping terminology---one concept can easily have four or five different names in the literature. Our ambition is to give an overview of this research area in different fields. We will not try to gather the theory into one unified framework. Instead, we hope that this review can help readers from one discipline to read and understand papers in others, aware of the confusing terminologies.

Another review of contributions primarily from computer science can be found in Santoro \textit{et al.}~\cite{santoro_etal} and an overview of contributions from the network engineering community can be found in Kuhn and Oshman~\cite{kuhn_oshman}.

In the rest of this paper, we will first discuss various real-world systems that can be modeled as temporal graphs. Then we go through theoretical developments, including measurements of temporal network structure, ways to meaningfully represent temporal networks as static networks, and studies of dynamical systems on temporal networks.

\section{Types of temporal networks}

\subsection{Person-to-person communication}

Records of electronic one-to-one communication are particularly suitable for the temporal network approach, especially in the context of the spreading dynamics of information or electronic viruses. Such data often come either in the form of lists of messages from one person to another at a point in time, or a dialogue between two persons within a time interval. The first type contains networks of e-mail messages~\cite{eckmann_etal,vazquez_etal,iribarren_moro,pan_saramaki}, mobile phone text messages~\cite{wu_etal,dan_etal}, and instant messages and messages in online forums~\cite{holme_2003,leskovec_horvitz}.  Phone calls are not instantaneous but have a specific duration, and can thus be considered to be of the second type~\cite{onnela_etal,candia_etal,karsai_etal,miritello_etal,pan_saramaki}. However, in many cases, call durations can be neglected and calls are assumed instantaneous. In this context, temporal network modeling and analysis of various temporal centrality measures (see Sect.~\ref{sec:centrality}) can be used for designing strategies for containing the spread of malware in mobile devices~\cite{tang_etal_2011}.

\subsection{One-to many information dissemination}

The broadcast of information to anyone that might listen, in contrast to one-to-one communication, is another type of information spreading between humans that could benefit from a temporal network approach. Typically, people have studied spreading events in blogs~\cite{kumar_etal,adar_adamic} or microblogs (like Twitter)~\cite{java_etal,kwak_etal}. Liben-Nowell and Kleinberg's study of chain-letter e-mails concerns an intermediate form of information transfer, between one-to-one and one-to-many~\cite{liben_nowell_kleinberg}. These studies have until now focused on aggregated statistics, without much focus on the temporal effects we discuss in this review (with Ref.~\cite{liben_nowell_kleinberg} as a bit of an exception, in that they also study response times), so further investigations are called for. Yasseri \emph{et al.}~\cite{yasseri11} take the time dimension into account in an interesting way in their analysis of the circadian patterns of Wikipedia editorial activity: such activity patterns can be used to estimate the geographical distribution of editors.

\subsection{Physical proximity}

Proximity patterns of humans---data on who is close to whom at what time---are important both for understanding the spread of airborne pathogens and word-of-mouth spreading of information. Such temporal networks have long been inaccessible for large-scale studies. Rather, researchers have performed tedious fieldwork in some confined space like a fraternity or an office~\cite{wasserman_faust}. Nowadays, it is fairly cheap to glean such information by using electronic devices. Here, the pioneering study was the Reality Mining project, where students of Massachusetts Institute of Technology were equipped with cell phones whose Bluetooth devices could detect their proximity to others~\cite{eagle_pentland}. The SocioPatterns project has developed a platform that allows physical proximity measurements based on wearable badges equipped with radiofrequency identification devices (RFID)~\cite{cattuto_etal}; these devices have been utilized in measurements of dynamic and temporal proximity networks of patients~\cite{Isella_PLoS1}, school children~\cite{Stehle2011PLos1}, and conference attendees~\cite{Stehle:2011nx}. Because the human body acts as a shield for the proximity-sensing RF signals, such sensors only record contacts when the individuals are facing each other, and thus a contact can also be considered as indicative of communication between the individuals~\cite{cattuto_etal,panisson_etal,isella_etal,stehle_etal_2011}. Recording of face-to-face communication events has also been realized with infrared sensor devices~\cite{takaguchi_etal}.
Another type of large-scale proximity data comes from hospitals where contacts between two patients that have been admitted to the same ward at the same time are recorded, sometimes including the medical staff~\cite{liljeros_etal_2007,ueno_masuda}. Such data is important for studying the dynamics of disease outbreaks~\cite{isella_etal,stehle_etal_2011} (such as MRSA) in hospitals and also protocols for wireless \textit{ad hoc} communication~\cite{panisson_etal}. Similarly, for livestock disease, it is important to know the movement of animals between farms etc.~\cite{paolo_etal,vernon_keeling}.

\subsection{Cell biology}

There are a handful of systems in cell and microbiology that can be modeled as networks~\cite{palsson}. Not all of these are dynamic enough in nature that one would benefit from modeling them as temporal networks. One of the systems that probably fits the framework is the interactome---the set of molecular interactions in a cell. The vertices of the interactome are proteins or lighter molecules that can attach or otherwise connect to one another to perform biological functions. Frequently, these interactions are represented as a static graph. However, much of the biological functionality comes from the fact that the connections are not active all the time. For this reason, Przytycka \textit{et al.}~\cite{przytycka_etal} believe that a ``shift from static to dynamic network analysis is essential for further understanding'' of the interactome. There is already a body of literature investigating the temporal aspects of protein interaction and gene-regulatory networks~\cite{taylor_etal,han_etal,komuro_white,lebre_etal}. This is perhaps the most natural level for temporal network approaches in cell biology---proteins are the workhorses of cells so any kind of cyclic, or otherwise dynamic, patterns have to be done by a shift in the interaction network. One can also represent gene expression and regulatory networks as temporal networks~\cite{lebre_etal,lebre,rao_hero_etal,yoshida_imoto_etal}. In these, the vertices are genes that  can be on (being transcribed) or not. Edges can be any of a number of functional relationships----that one gene (via feedback from RNAs or proteins) affect the transcription of another, or that two genes code for proteins that interact, or that they are close to each other on the DNA, etc.

Another biological network that changes over time is the metabolism---the set of chemical reactions that occur in a healthy organism. The vertices in metabolic networks are molecular species that are connected if they are involved in the same chemical reaction. At any given time and subcellular localization, only a part of the entire biochemical reaction system is active. This situation changes with time, and temporal networks can potentially capture its dynamics~\cite{chechik_etal}. However, the change comes about via alterations in the influx to the cell that reflects the overall state of the organism's body, or is controlled by genes. Both these processes are relatively slow compared to the conversion of molecules, so probably a temporal-network analysis of metabolism does not need the more elaborate methods that we mention in this review.

\subsection{Distributed computing}

Much of the early theoretical developments on temporal networks come from computer science. There are many different types of distributed computing systems but they all consist of fairly independent computational units spread out over some network~\cite{ghosh}. Since the computation runs in parallel to the information spreading between the units, they typically need to operate with information that is of different age . To study such a system theoretically, a central problem is estimating and controlling the age of the information that is accessible to the vertices.

\subsection{Infrastructural networks}

Most infrastructural networks change so slowly that there is no point in modeling them as temporal networks. Take the Internet as an example: the dynamical system in question---the flow of data packets---operates globally at a time scale of seconds. The fastest changes to the network topology come from new business agreements between subnetworks that are already in physical contact. These happen globally a few times per minute, but compared to the size of the entire Internet, this is so slow that one can probably safely assume it is static~\cite{farrel,pastor_santorras_vespignani_2004}. However, for some types of transport networks, e.g.\ the air-transport network, it can be meaningful to apply certain temporal network concepts such as temporal path durations and centrality~\cite{pan_saramaki}. In this case, edge activation sequences correspond to scheduled transport connecting vertices, such as individual flights or trains.

\subsection{Neural and brain networks}

Networks of neural connections represent another class of biological networks that may benefit from the temporal network approach. There are several levels of structural and temporal connectivity, from the spiking patterns of individual neurons to more coarse-grained physiological or functional connections between brain areas~\cite{sporns_etal,bullmore_sporns}. For the latter, there are different experimental modalities with different tradeoffs regarding spatial or temporal resolution. Electroencephalography (EEG) and magnetoencephalography (MEG) measure electrical signals and perturbations of the extracranial magnetic fields, respectively, and have fairly good temporal but poor spatial resolutions. Functional magnetic resonance imaging (fMRI) detects changes in regional brain activity by measuring blood oxygenation levels, and it has a high spatial but poor temporal resolution. Regardless of the experimental technique used, the typical approach is to use time series associated to vertices (individual sensors for EEG and MEG, three dimensional regions---voxels---or larger aggregated regions for fMRI), and assign an edge between two vertices at a given point in time or within some time window, if the signals are correlated or in phase. Such networks are called functional networks; in this abstraction, the existence of a link represents simultaneous activation of brain areas, indicating a functional connection between them (see, \emph{e.g.}~\cite{bullmore_sporns}). Naturally, such functional connections reflect the properties of the underlying anatomical connectivity network between brain areas  mediated via neuronal fiber bundles (the structural network), but functional links may appear between brain areas that have no (or only few) direct physical connections. In general, in temporal brain functional networks, the temporal links represent  the time dynamics of simultaneous brain area activations  -- while the structural substrate network is static on such time scales, functional link activations vary in time. 

As an example of a static approach to functional brain networks, De Vico Fallani \textit{et al.}~\cite{devicofallani_etal} use correlations of EEG time series to derive a directed network in which they analyze the motifs (over-represented subgraphs of three vertices). Such a study would be even richer in the temporal network framework where the motifs would represent temporal sub-networks. 
There are, to our knowledge, only a few papers where the time domain is directly taken into account: Valencia \textit{et al.}~\cite{valencia_etal} study functional brain networks reconstructed from MEG data with the phase-locking criterion, and show that the functional connectivity varies with time and frequency during the processing of visual stimuli, while certain network features such as small-world characteristics are maintained (see also \cite{tang_etal_2010}). Dimitriadis \textit{et al.}~\cite{dimitriadis_etal} investigate brain dynamics as measured with EEG during mental calculations, and identify hubs that facilitate communication in the underlying functional networks. Bassett \textit{et al.}~\cite{bassett_etal} monitor the evolution of a brain network while the subject is learning a simple motor task. In addition, it would be of great interest to measure the dynamics of functional networks when the applied stimulus is also time-dependent, especially with naturalistic (close-to-real-life) paradigms such as watching a movie or listening to music in the fMRI scanner (see e.g.\ Ref.~\cite{kauppi_etal}).

\subsection{Ecological networks}

\begin{figure}
\includegraphics[width=\linewidth]{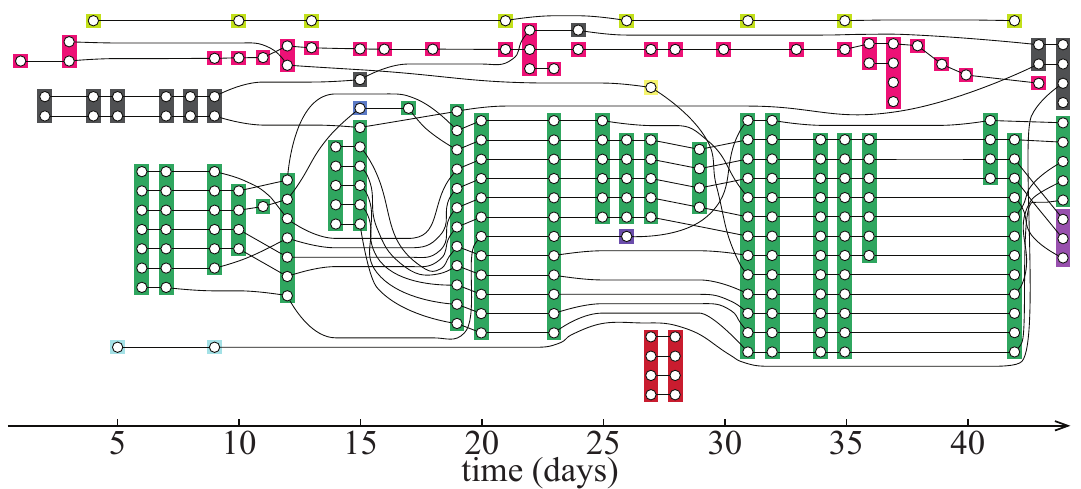}
\caption{A temporal network of zebras. The figure is adapted from Tantipathananandh \textit{et al.}~\cite{tantipathananandh_etal}. The data comes from Sundaresan \textit{et al.}~\cite{sundaresan_etal}. Each horizontal line corresponds to one individual. The contacts between individuals are not shown; instead the clusters as identified by the algorithm in Tantipathananandh \textit{et al.}\ are illustrated by the colored squares.}
\label{fig:tantipathananandh}
\end{figure}

Ecological networks capture the interactions between species or other categories of organisms~\cite{sole_bascompte,pascual_dunne}. They could be trophic, i.e.\ showing which species prey on another, or mutualistic, i.e.\ representing how two species engage in a relationship that benefits both. In many aspects, ecological networks are dynamic~\cite{deruiter_etal}. As an example, they can change with the seasons as organisms go through different phases of their life cycles (and thus have different capabilities and needs with respect to their interaction with others)~\cite{pahlwostl,ulanovicz}. As another example, at longer time scales, ecological networks change through evolution. Since most methods of this review are specialized to cases where the time scale of topological changes of the network is not too much slower than the dynamics of the network (the flux of matter between species), studies of evolutionary effects might not require the temporal network approach (at least in their traditional sense, cf.\ Ref.~\cite{yoshida_etal}). For faster changes of the interaction patterns, in response to environmental changes, yearly and circadian cycles, etc., temporal networks could provide a useful framework.

In population biology one also studies proximity and mobility networks of animals~\cite{lusseau_etal,croft_etal,sundaresan_etal,tantipathananandh_etal,blonder_dornhaus}; see Fig.~\ref{fig:tantipathananandh}. These are, just like human proximity networks, prime examples of systems where the temporal dimension can affect dynamical systems like disease and information spreading, and are thus apt for temporal-network analysis. In Ref.~\cite{bajardi2011}, dynamical patterns of cattle movement were analyzed with temporal networks where vertices represent premises, and edges cattle movement among premises.

\subsection{Other systems}

The above-mentioned systems are far from the only potential applications of temporal network modeling. Probably, the easiest way of finding more examples is to look at the complex-network literature and to ask oneself if a certain system has enough temporal structure for a temporal-network approach. An early paper on time-evolving network considered supply networks for the manufacturing industry~\cite{dagum_etal}. There are likely other economic systems that would benefit from temporal network modeling. Networks that, like citation networks~\cite{medo} are normally thought of as strictly growing, could show temporal effects in the growth that could benefit from being studied in a temporal-network framework.

\begin{figure}
\includegraphics[width=0.8\linewidth]{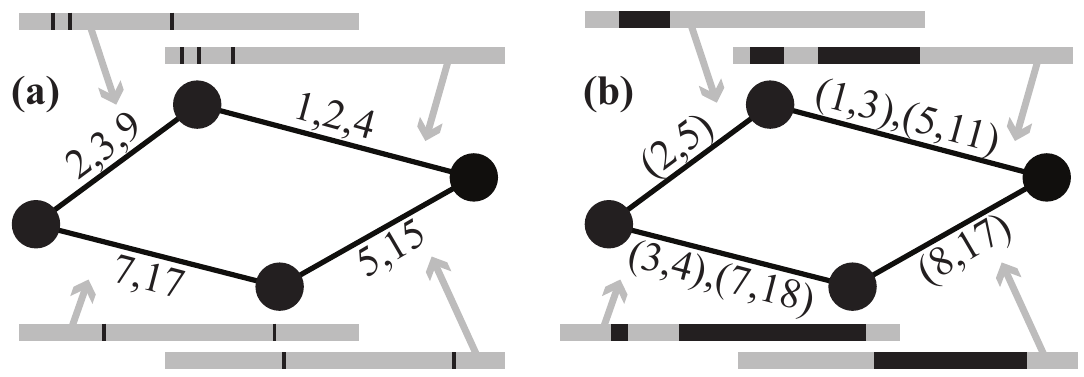}
\caption{Contact sequences and interval graphs. This figure illustrates the two fundamental temporal network representations in our discussion---contact sequences (a) and interval graphs (b). The times of the contacts are states next to the edges. We also visualize the contacts timelines (grey bars). In these the contacts are marked by black bars or fields and the time lines range from $t=0$ to $t=20$ (with $t=0$ to the left). In the former, contacts occur at points in time whereas the contacts are extended in time in the latter. Timeline plots like those in Figs.~\ref{fig:reachability}(b) and \ref{fig:limits_of_temporal_models}(b) are another suitable depiction of contact sequences, for contact graphs such illustrations are not as readable.}
\label{fig:types}
\end{figure}

\section{Preliminaries}

The temporal networks we consider in this review can be divided into two (rough and overlapping) classes corresponding to the two types of representations illustrated in Fig.~\ref{fig:types}. In the first representation (Fig.~\ref{fig:types}(a)) there is a set of $N$ vertices $V$ interacting with each other at certain times, and the durations of the interactions are negligible. In this case, the system can be represented by a \emph{contact sequence}---a set of $C$ \emph{contacts}, triples $(i,j,t)$ where $i,j\in V$ and $t$ denotes time. Equivalently, one can represent the system by $V$, a set of $M$ edges (pairs of vertices) $E$, and, for $e\in E$, a non-empty set of times of contacts $T_e=\{t_1,\dots,t_n\}$. Typical systems suitable to be represented as a contact sequence include communication data (sets of e-mails, phone calls, text messages, etc.), and physical proximity data where the duration of the contact is less important (e.g.\ sexual networks). Commonly, authors group the contacts happening at the same discrete timestep into one graph (or ``graphlet'' in the terminology of Ref.~\cite{basu_etal}) and present the temporal network as a time sequence of graphs. Since this representation makes it tempting to think of the temporal-network structure as an evolving static network structure (which misses many of the unique points of temporal networks), we prefer contact sequences. Furthermore, in many real datasets with a high time resolution there are only a handful of edges present at a timestep among tens of thousands of vertices which make the graph-sequence representation a little odd (but this is really just a matter of taste).

In the second class of temporal networks we discuss, \emph{interval graphs}, the edges are not active over a set of times but rather over a set of intervals $T_e=\{(t_1,t'_1),\dots(t_n,t'_n)\}$, where the parentheses indicate the periods of activity---the unprimed times mark the beginning of the interval and the primed quantities mark the end. The static graph with an edge between $i$ and $j$ if and only if there is a contact between $i$ and $j$ is called the \emph{(time) aggregated graph}. Examples of systems that are natural to model as interval graphs include proximity networks (where a contact can represent that two individuals have been close to each other for some extent of time), seasonal food webs where a time interval represents that one species is the main food source of another at some time of the year, and infrastructural systems like the Internet. Like for static graphs, it can be useful to define an index function of whether a pair of vertices is connected at a given time. This is the \emph{adjacency index} (Ref.~\cite{casteigts_etal} call it ``presence function'')
\begin{equation}\label{eq:adjacency}
a(i,j,t)= \left\{\begin{array}{ll}1 & \mbox{if $i$ and $j$ are connected at time $t$}\\0 & \mbox{otherwise}\end{array}\right .
\end{equation}
Just as the largest body of literature on network theory considers simple graphs (of undirected edges that never occur twice between the same vertices, and never connect a vertex to itself), we put (unless otherwise stated) some further restrictions on the time stamps of the edges. We assume that a triple of a contact sequence never occurs twice, which means that we can order the contacts uniquely (first by the time stamps, then by their smallest vertex index and finally by their largest vertex index). Usually, we also disregard the order of the vertices in a contact---if contacts are considered directed, we will always indicate this separately. For interval graphs, we assume that there are no empty or overlapping intervals. More mathematically speaking, consider two intervals $(t_i,t'_i),(t_j,t'_j)\in T_e$; then the following three statements need to be true
\begin{enumerate}
\item $t_i<t'_i$
\item $t_j<t'_j$
\item $t_i<t_j$ if and only if $t'_i<t_j$
\end{enumerate}
These definitions can of course be extended in many ways---one can think of weighted temporal networks (where a vertex or edge is associated with a time-dependent scalar), networks where the edges take some time to traverse~\cite{buixuan_etal} or the contacts are completed only after some duration $\delta t$ and should thus be represented as quadruples $(i,j,t,\delta t)$ \cite{pan_saramaki}. Of course, vertices could also be active intermittently, but usually this is reflected in the activity of edges and we will not discuss this issue further. Such extensions might require modifications of the methods and measures we discuss in this review. Some such modifications are straightforward, while others are open research questions; several concepts of temporal graphs have no immediate counterpart in static graphs. 

There are other representations present in the literature. Those that reduce the information from the original temporal graph by mapping them to a static graph are discussed in Sect.~\ref{sec:static}. A yet rarely followed path is Harary and Gupta's suggestion to model temporal graphs with logic programming~\cite{harary_gupta}.

\section{Measures of temporal-topological structure}

\subsection{Introduction}

The topological structure of static networks can be characterized by an abundance of measures (see, e.g.,~\cite{daCostaSurvey}). In essence, such measures are based on connections between neighboring nodes (such as the degree or clustering coefficient), or between larger sets of nodes (such as path lengths, network diameter and various centrality measures). When the additional degree of freedom of time is included in the network picture, many of these measures need rethinking or revising. While some measures are perhaps best applied to networks aggregated over chosen time periods (\emph{e.g.} the time-dependent degree of a node can be computed as the number of links activated within some time window), other properties are directly influenced by the order of link activations. As an example, \emph{paths} that transmit anything through the network need to follow time-ordered sequences of contacts, and like the temporal networks themselves, such paths are not static but change in time. In this section, we will review measures proposed for characterizing temporal-topological structure. Many of these build on the concept of time-respecting paths discussed in the first subsection, such as different centrality measures. We also address some of the few methods proposed for characterizing mesoscopic features and patterns in temporal networks; in this area, there is still a clear lack of methods. We conclude by discussing measures of the temporal inhomogeneities of contact sequences and information-theoretic aspects.

\subsection{Time-respecting paths and reachability}

Paths that connect nodes represent the pathways constraining the dynamics of any process taking place on the network. In a static graph, a path is simply a sequence of edges such that one edge ends at the node where the next edge of the path begins (such as A to B to C to D in Fig.~\ref{fig:reachability}). 
In order for this concept to be meaningful in temporal networks, especially in relation to dynamical processes, paths must necessarily be constrained to
sequences of link activations that follow one another in time. Thus, in a temporal graph, paths are usually defined as sequences of contacts with non-decreasing times that connect sets of vertices. Kempe \textit{et al.}~\cite{kempe_etal} and other authors~\cite{holme_etal} call such paths ``time-respecting.'' As an example,
in Fig.~\ref{fig:reachability}, there are time-respecting paths from A to D  (for example ${(\mathrm{A},\mathrm{B},7), (\mathrm{B},\mathrm{C},8), (\mathrm{C},\mathrm{D},11)})$ but none from A to E. In the literature, the terms ``journey''~\cite{buixuan_etal,ferreira} and ``non-decreasing path''~\cite{cheng_etal} have also been
used for time-respecting paths. 

The constraint of having to follow time-ordered sequences of contacts gives rise to differences between temporal paths and paths in static networks.
Similarly to static directed networks, it might be the case that $i$ is reachable by time-respecting paths from $j$, but $j$ cannot be reached from $i$. A difference between directed and temporal networks is that the paths are not transitive. The existence of time-respecting paths from $i$ to $j$ and $j$ to $k$ does not imply that there is a path from $i$ to $k$, as seen in the example above -- a path from $i$ to $k$ via $j$ exists only if the first contact on the $j-k$ path takes place after the last contact on the $i-j$ path. This is related to a fundamental property of time-respecting paths: they, too, are temporal, and begin and end at certain points in time. Thus, the existence of a time-respecting path that begins at $i$ at time $t'$ and leads to $j$ does not guarantee
that such a path between $i$ and $j$ exists for $t>t'$; in addition, a future temporal path joining $i$ and $j$ might follow a different route. Hence, the statement that "there is a time-respecting path between $i$ and $j$" is ambiguous; such paths always take place within some time window.

Thus, time-respecting paths define which vertices can be reached from which other vertices within some observation window $t\in[t_0,T]$.
The set of vertices that can be reached by time-respecting paths from vertex $i$ is called the \emph{set of influence} of $i$. This is important \emph{e.g.}~for disease spreading, as it is the set of vertices that can eventually be infected if $i$ is the source of infection. It may be useful to define a set of influence at the time $t$ as the set of vertices that can be reached via time-respecting paths from vertex $i$ that begin at time $t$ or later. Holme~\cite{holme_2005} calls the average fraction of vertices in the sets of influence of all vertices as the \emph{reachability ratio}. 

Reversely, one can also define the \emph{source set} of $i$ as the set of vertices that can reach $i$ through time-respecting paths within the observation window. This set consists of all vertices that can have been the source of an infection infecting $i$. Riolo \textit{et al.}~\cite{riolo_etal} points out the size of $i$'s source set---$i$'s \emph{source count}---as an important quantity. Moody~\cite{moody} gives another definition of the reachability of a vertex that increases in proportion to the count of time respecting paths.
Again, as the source set is time-dependent, one may also monitor the source count a function of time, \emph{i.e.} study how many other vertices may reach vertex $i$ by time-respecting paths by time $t'$, when the paths begin no earlier than $t<t'$. It may be useful to view the two time-dependent sets -- the source set and the set of influence -- as the past and future "light cones" for vertex $i$, \emph{i.e.}~the set of nodes which may have influenced $i$'s current state and the set of nodes which may be influenced by $i$ in the future via time-respecting paths (see also Sect.~\ref{sec:components} below).

\subsection{Time-respecting paths with limits on waiting times}
\begin{figure}
\includegraphics[width=0.9\linewidth]{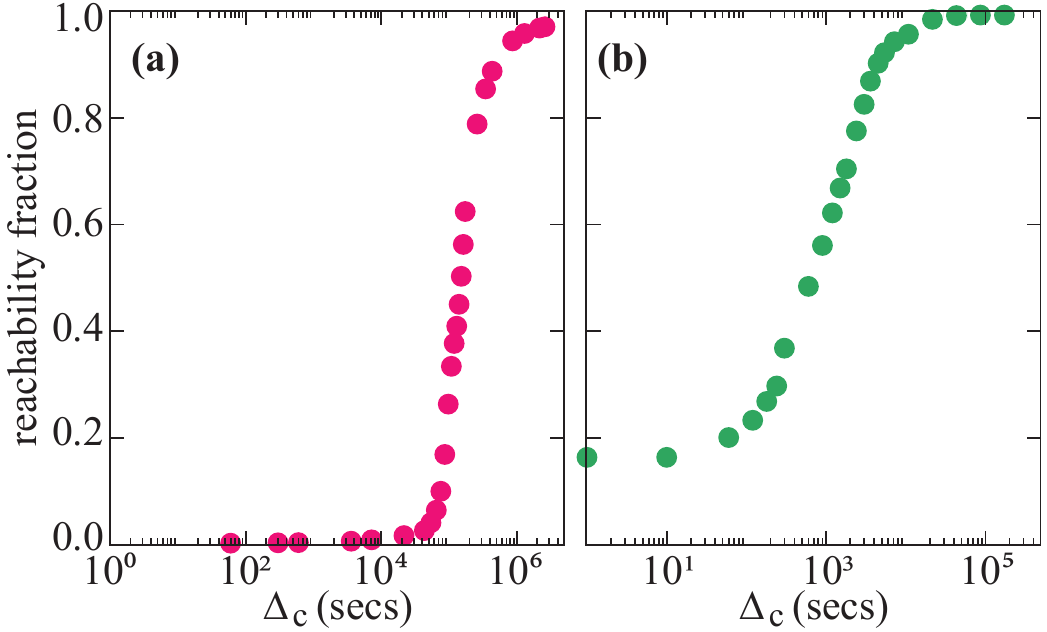}
\caption{The reachability ratio as a function of the maximum allowed delay of a time-respecting path at each vertex. Panel (a) shows data for a mobile telephone call network, (b) comes from the connections of an airline system. This figure is adapted from Ref.~\cite{pan_saramaki}.}
\label{fig:pan_saramaki}
\end{figure}

Pan and Saram\"aki~\cite{pan_saramaki} note that some spreading or transport processes that follow time-respecting paths set limitations on the times that the paths are allowed to spend at vertices, i.e.\ times between two consecutive contacts on a path. These limitations may be from below---such that the process must be allowed to wait for some time before the next contact on the path---or from above, as for spreading dynamics, where the transmission has to happen quickly enough before infectious nodes recover, or before nodes lose their interest in forwarding some piece of information. They set a limit  for the latter---the maximum allowed waiting time at a vertex---and measure the reachability ratio as a function of empirical networks of cell-phone calls and passenger flights (see Fig.~\ref{fig:pan_saramaki}). The reachability ratio was observed to increase rather sharply around a characteristic time of about two days for the cell-phone data and 30 minutes for the airline network. These time scales can be interpreted as reflecting some fundamental property of the system. Mobile phone call sequences are known to be bursty and this is reflected in long inter-call intervals, and thus information must still be further transmitted two days after its reception if it is to reach a large number of individuals. Interestingly, 30 minutes is about the minimum allowed transfer time between flights for transferring passengers.

\subsection{Connectivity and components}\label{sec:components}

Connectivity---whether or not a pair of vertices is connected by a path---is a fundamental concept for networks. Any network can be divided into sets of nodes based on their connectivity; these sets, in turn, impose limitations on any dynamics taking place on the network. For static networks, vertices are either connected or not, and connected components are defined as sets of vertices between which some path can always be found. As mentioned above, connectivity is not a symmetric relation for directed or temporal graphs. In directed graphs, the property of connectivity can be divided into two parts: \emph{strong connectivity}, where there is a directed path between all pairs of vertices, and \emph{weak connectivity}, where there is a path between all pairs of vertices if the edges are considered undirected. These two concepts can be generalized for temporal networks. Nicosia \textit{et al.}~\cite{nicosia_etal} propose the following definitions: two vertices $i$ and $j$ of a temporal network are defined to be \emph{strongly connected} if there is a directed, time-respecting path connecting $i$ to $j$ and vice versa, while they are \emph{weakly connected} if there are undirected time-respecting paths from $i$ to $j$ and $j$ to $i$, i.e.\ the directions of the contacts are not taken into account. On the basis of these definitions, one may then define strongly or weakly connected components of the temporal graph as sets of vertices where each pair fulfills these criteria. Nicosia \textit{et al.}\  also show that the problem of finding strongly connected components in temporal graphs can be mapped into the problem of finding maximal cliques in affine graphs, where an element of the adjacency matrix indicates strong connectivity between the respective vertices in the temporal graph. One should keep in mind that such properties always depend on the time of measurement, i.e.\ are in general only valid within some specified time window. In addition to strong and weak connectivity, one can define yet another type of connectivity---transitive connectivity---for temporal networks. A subgraph is transitively connected if time respecting paths from $i$ to $j$ and $j$ to $k$ implies a time respecting path from $i$ to $k$.

\subsection{Distances, latencies, and fastest paths}

For static networks, the \emph{geodesic distance} between two vertices is defined as the length of the \emph{shortest path} joining them, path length being defined as the number of links forming a path. Shortest path lengths obviously influence how quickly anything can propagate between nodes, and their average and distribution determines the overall "compactness" of a network. Evidently, when the dimension of time is added to the picture, it is useful to define similar quantities characterizing how \emph{quickly} vertices can reach each other through time-respecting paths; here, to the best of our knowledge, the earliest work is by Cooke and Halsey in the 60's~\cite{cookehalsey}. Here, some difficulties arise because the time-respecting paths are themselves temporal and because the observation window is always finite, and  choices have to be made \emph{e.g.}~regarding proper ways of averaging over quantities. In addition, the nomenclature in the literature has not yet converged.

Obviously, any time-respecting path is associated with a \emph{duration}, measured as the time difference between the last and first contacts on the path; note that some authors have called it the \emph{temporal path length}~\cite{pan_saramaki}. Analogously to the shortest paths that define the geodesic distance, one can find the \emph{fastest} time-respecting path(s) between two nodes; the shortest time within which $i$ can reach $j$ is called their \emph{latency} (also "temporal distance" \cite{pan_saramaki}). As the concepts of temporal duration and link-wise distance have been used interchangeably in the literature, we will in the following reserve the word "distance" for measuring numbers of links, and "duration" and "latency" for measuring times (see also Sect.~\ref{sec:average_latency} below).

The concept of latency was originally introduced in the study of distributed computation. A central problem in this area is keeping track of the age of information that a vertex has about other vertices. A quite reasonable assumption is that vertices in contact update each other's information so that after a contact, both vertices share the most recent information that either of them had before the contact. This scenario is similar to the SI disease spreading model that we will discuss later, if the disease transmission probability upon contact is set to 100$\%$. In terms of information spreading, this contact process defines the fastest possible trajectories of information between vertices.

Consider the vertex $i$ at time $t$ in a temporal network over which information spreads. Then let $\phi_{i,t}(j)$ denote the latest time before $t$ such that information from $j$ can have reached $i$ by time $t$. We call this quantity $i$'s \emph{view} of $j$'s information at time $t$. Furthermore, $\lambda_{i,t}(j) = t-\phi_{i,t}(j)$ is called $j$'s \emph{information latency}, or just \emph{latency}, with respect to $i$ at time $t$, and is thus a measure of how old $i$'s information coming from $j$ is at time $t$. Finally, the vector $[\phi_{i,t}(1),\dots,\phi_{i,t}(N)]$ is called $i$'s \emph{vector clock}. This framework was introduced by Lamport~\cite{lamport} and further developed by Mattern~\cite{mattern}. Note that the above definition looks backwards in time; one may also define a forward latency (called ``temporal distance'' in Ref.~\cite{pan_saramaki}) $\tau_{ij}(t)$ that measures how long it takes to reach $j$ from $i$ along the fastest path, when the measurement begins at time $t$. A bit reminiscent to vector clocks, Panisson \textit{et al.}~\cite{panisson_etal} introduce what they call \textit{intrinsic time} to denote the active time of every vertex. This, they argue, is useful to study the temporal statistics of information spreading.


\begin{figure}
\includegraphics[width=0.8\linewidth]{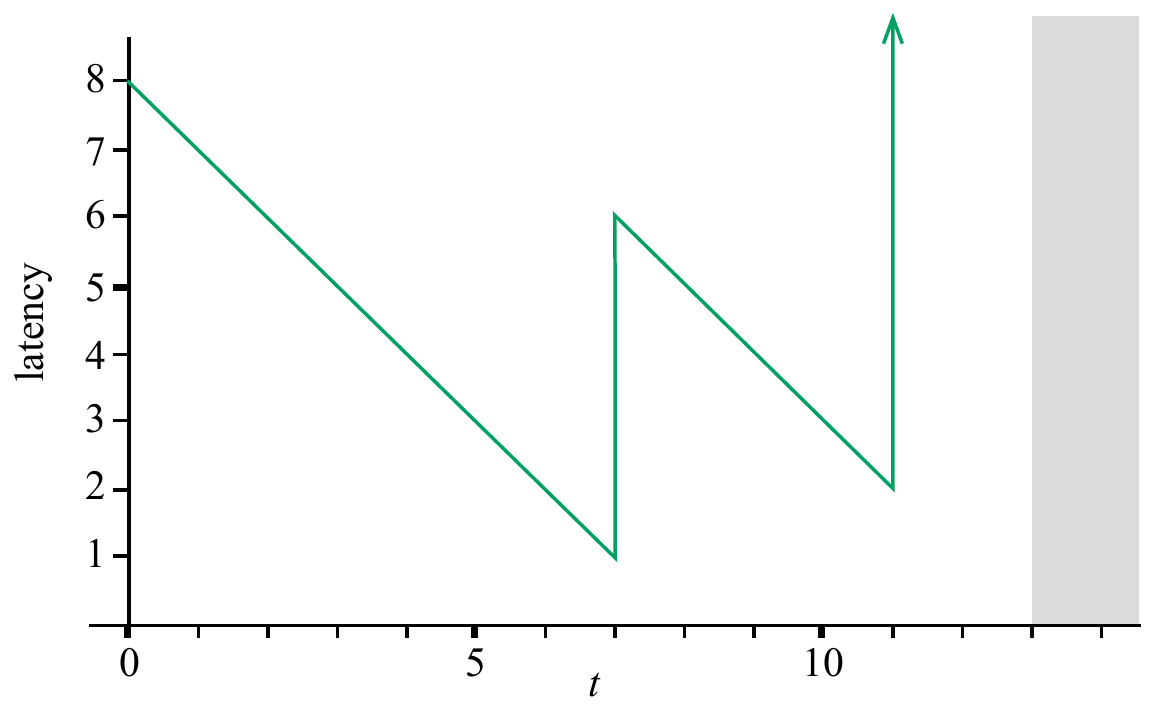}
\caption{The forward latency for paths from A to C of Fig.~\ref{fig:reachability} as a function time. There are two paths joining A and C that go through an arbitrary set of vertices---the first contact of the first path takes place at $t=7$, and the duration of the path is one unit of time, i.e.\ the path arrives at $C$ in one time unit. The next path begins at $t=11$, and takes two units of time to traverse. If one would use periodic temporal boundary conditions for paths between A and C, so that the first observed path joining them at $t=7$ repeats at $t=T+7$ where $T=13$ is the observation period limit, then the arrow would go up to latency $10$ and then fall down linearly like for early times and thereby repeat a cyclic pattern. This figure is adapted from Ref.~\cite{pan_saramaki}.}
\label{fig:sawtooth}
\end{figure}


\subsection{Average latency}
\label{sec:average_latency}

A natural use for durations and latencies is in characterizing the overall ``velocity'' of the temporal network, i.e.\ measuring how quickly vertices can on average transmit something to each other along the contact sequences. However, taking an average over the entire window of observation to get a value for the entire graph---or even only for a pair of vertices---is not that straightforward. The problem is related to the finiteness of the observation period, and the fact that there are typically different time-respecting paths between the same pair of vertices that begin at different points in time. Furthermore, close to the end of the observation window, time-respecting paths become rare as they do not have enough time to be completed, and vertices may no longer reach each other. One possible quantity for measuring the velocity of paths in general is to enumerate all fastest time-respecting paths between vertices and then compute the average duration of such paths. However, this measure would not reflect the frequency of the paths, and would not be affected by waiting times before the first contacts of such paths: if $i$ and $k$ are joined by one single path that takes one unit of time to traverse or by ten paths of the same duration, this average would equal unity in both cases.

For the average latency, measuring how long it on average takes to reach $i$ from $j$, the situation is more difficult. Latency varies with time with a saw-tooth pattern (cf.\ Fig.~\ref{fig:sawtooth}), where the jumps occur at points where a new fastest time-respecting path begins. Close to the end of the observation window, latency becomes infinite, as paths no longer have enough time to be completed. Averaging only over the period of finite latency is problematic: if vertices $i$ and $j$ were connected only by a single path that takes place late in the observation period, their average latency would be high, whereas it would be low if that path took place earlier. To account for this, Pan and Saram\"aki~\cite{pan_saramaki} proposed a pair-specific temporal boundary condition, where for every pair of vertices, the first observed path joining them is repeated once when computing the latency for that specific pair. Another possibility would be to periodically repeat the entire temporal contact sequence as was done e.g.~in Ref.~\cite{karsai_etal}. However, this procedure may give rise to artifacts and connect pairs of vertices that are not connected at all within the observation window.

For long enough periods of observation, another difficulty is posed by the dynamics of vertices entering and exiting the system. If only edge activation sequences are observed, such vertex dynamics cannot generally be distinguished from edge dynamics---e.g.\ in a temporal network spanned by telephone calls, even if a person makes only infrequent calls, one cannot generally assume that the person wasn't a subscriber to the operator before the first call, or has left the operator after the last call. However, in some cases external information on vertices is available, and then one may choose to only include vertices that are known to be part of the system for the entire period.

Finally, as a word of caution about the nomenclature, as already mentioned above, some authors use the terms ``distance'' and ``length'' as measures of time---e.g.\ Kossinets \textit{et al.}~\cite{kossinets_etal} define the ``distance'' between two vertices as the shortest duration of any time-respecting path between them. Tang \textit{et al.}~\cite{tang_etal_2009} calls the average time to reach (the reachable) vertices for time-respecting paths starting in a time window early in the data ``temporal path length''. To be fair, we should point out that ``distance'', as in the standard static graph definition is, being a dimensionless quantity, a bit of a misnomer too. Furthermore, what we call average latency has several names: ``reachability time'' in Ref.~\cite{holme_2005}, ``temporal proximity'' in Kostakos~\cite{kostakos}~\footnote{Kostakos defines a set of measures for generation graphs, in increasing specificity (some corresponding to latency, some corresponding to the shortest latency over a time interval) which he all calls temporal proximity.}, ``characteristic temporal path length'' in Tang \textit{et al.}~\cite{tang_etal_2009} and ``temporal distance'' in Ref.~\cite{pan_saramaki}.

\begin{figure}
\includegraphics[width=\linewidth]{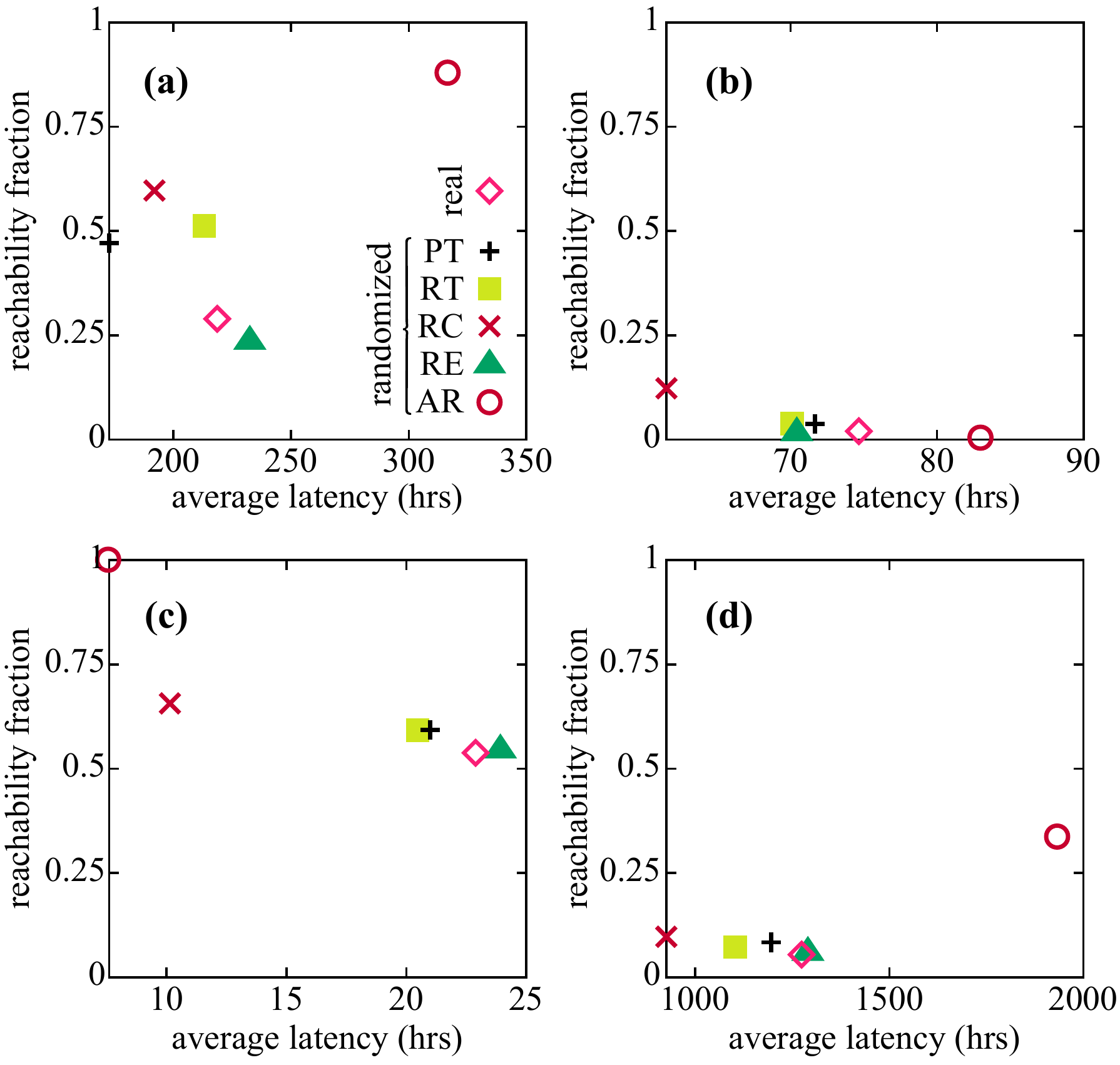}
\caption{Average latency and reachability ratio of some empirical contact sequences. This figure is reprinted from Ref.~\cite{holme_2005}. Each panel corresponds to a dataset: (a) is from contacts of an Internet community; (b)--(d) comes from e-mail exchange, for details see Holme~\cite{holme_2005}. The randomizations are Permuted Times (PT), Random Times (RT), Random Contacts (RC), Randomized Edges (RE) and All Random (AR).}
\label{fig:holme}
\end{figure}

\subsection{Diameter, network efficiency}

There are several quantities that characterize the compactness of a static network in terms of path lengths -- in addition to average shortest path lengths, the largest distance between any pairs of vertices defines the \emph{diameter} of the network, and the average over inverse path lengths of all paths the \emph{efficiency}.
One option of defining the diameter in a temporal graph would be to take the longest average latency (although one could again argue that one should avoid the mix-up between quantities whose names relate to length but that are measured in units of time). Again, one would then have to choose how to deal with infinite latencies, i.e.\ pairs of vertices that are not connected by any time-ordered path within the observation window. Chaintreau \textit{et al.}~\cite{chaintreau_etal} gives another definition by requiring that the diameter should be a number as small as possible such that almost surely, increasing it would not make you find more pairs of vertices connected by a time-respecting path that is shorter than the diameter. The advantage with this definition is that it does not require all vertex pairs to be connected, which is usually the case for empirical data sets.

Tang \textit{et al.}~\cite{tang_etal_2009} propose \emph{network efficiency} (the harmonic average of the latency) as a distance metric for temporal networks. This measure combines the average latency and reachability ratio of Holme~\cite{holme_2005} by a harmonic mean. It was also mentioned in Holme~\cite{holme_2005} but discarded with the argument that the two properties (of how many vertex pairs that are connected by time-respecting paths and how fast information can reach between them) carry very different information about the function of the system and should rather not be mixed up (see Fig.~\ref{fig:holme}).

\subsection{Minimum spanning tree}
The minimum spanning tree is an important concept related to paths in static weighted graphs. It is defined as a subgraph that is a tree of minimal total weight that connects all the vertices of the graph.  Such minimum spanning trees are important for engineering applications as it they the cheapest way to reach all vertices, when the cost of an edge is measured in terms of weight. Gunturi \textit{et al.}~\cite{gunturi_etal} presents a way to identify a related concept in temporal networks that they call ``time-sub-interval minimum spanning tree''. This is roughly speaking the spanning tree in an interval that has the smallest average latency.

As an unusual application of reachability analyses we mention Ref.~\cite{kuwata_etal} that uses a temporal network approach to study how to navigate hot-air balloons over a planet with a predictably changing wind field. In this case the vertices are cuboid volumes of the atmosphere connected if a balloon has the possibility of moving from one to another. (Normally, one probably needs more complex topologies to benefit from a temporal-network approach.)

\subsection{Centrality measures}
\label{sec:centrality}

In network theory, numerous centrality measures have been defined for identifying important vertices beyond the degree, \emph{e.g.} with respect to their average distance to other vertices or importance for shortest paths connecting other vertices. As when translating other quantities from static to temporal graphs, there is usually no unique candidate for the temporal version of a static centrality measure. A rather straightforward approach, however, is to replace the role of paths in static networks by time-respecting paths. In the words of Moody~\cite{moody}: ``time-dependent centrality measures that make use of the number and length of time-[respecting] paths are obvious choices.'' The \emph{closeness centrality} $C_C$~\cite{newman_2010} is for static networks defined as
\begin{equation}\label{eq:closeness}
C_C(i)=\frac{N-1}{\sum_{j\neq i} d(i,j)},
\end{equation}
where $d(i,j)$ is the geodesic distance between $i$ and $j$, \emph{i.e.}~the closeness centrality measures the inverse total distance to all other vertices and is high for vertices who are close to all others. Similarly, for temporal networks, one may be interested in how quickly a vertex may on average reach other vertices, and define the temporal closeness centrality as~\cite{tang_etal_2010}
\begin{equation}\label{eq:tempo_closeness}
C_C(i,t)=\frac{N-1}{\sum_{j\neq i} \lambda_{i,t}(j)},
\end{equation}
where $\lambda_{i,t}(j)$ is the latency between $i$ and $j$. Just like the static closeness centrality is undefined for disconnected graphs, the temporal version does not work unless there is a time-respecting path from $j$ to $i$ ending at time $t$ the latest. This is, in practice, quite a hard restriction for temporal networks, at least if $t$ is rather late in the observation period. In some situations one may want to get rid of the time dependence by averaging it out in Eq.~(\ref{eq:tempo_closeness}). In this case one would have to choose $t$ in the interval when $\lambda_{i,t}(j)$ is finite for all $i$ and $j$, and integrate the sawtooth-like latency function (see Fig.~\ref{fig:sawtooth}) in this interval, or to apply the boundary-condition based method introduced in Ref.~\cite{pan_saramaki}. The downside of this approach is that there is potentially interesting information in the intervals with infinite latencies, and it may therefore be better to use other centrality metrics. An augmented closeness centrality measure based on reciprocal latencies in the same vein as the above-mentioned ``efficiency''  can be defined as
\begin{equation}\label{eq:tempo_efficiency}
C_E(i,t)=\frac{1}{N-1}\sum_{j\neq i} \frac{1}{\lambda_{i,t}(j)}
\end{equation}
where $\frac{1}{\lambda_{i,t}(j)}$ is defined as zero if there are no time-respecting paths from $j$ to $i$ arriving at time $t$ or earlier~\cite{santoro_etal,pan_saramaki}. A similar definition can be used for closeness centrality based on pairwise latencies averaged over the observation window; if there are no paths at all between $i$ and $j$, the average latency is infinite for this pair and the reciprocal latency is defined as zero~\cite{pan_saramaki}.

 The \emph{betweenness centrality} $C_B$~\cite{jackson,easley_kleinberg} is another important centrality measure based on shortest paths, measuring the fraction of shortest paths passing through the focal vertex (or edge). For static networks, betweenness centrality is formally defined as
 \begin{equation}\label{eq:betweenness}
C_B(i)=\frac{\sum_{i\neq j\neq k}\nu_i(j,k)}{\sum_{i\neq j\neq k}\nu(j,k)}
\end{equation}
where $\nu_i(j,k)$ is the number of shortest paths between $j$ and $k$ that pass $i$, and $\nu(j,k)$ is the total number of shortest paths between $j$ and $k$. This definition is straightforwardly generalizable~\cite{tang_etal_2010b}
to temporal networks by adding a dependence on time $t$ and counting the fraction of shortest or 
fastest time-respecting paths that pass through the focal vertex. In the first alternative, the focus is on paths that contain the smallest number of contacts, and in the 
second, on paths that are fastest to traverse.
Here, one also has
to define the temporal boundaries---options are e.g.\ to use the whole observation period, or only consider the fastest time-respecting paths that begin at times closest to $t$ at each vertex $j$. Just like the regular betweenness, the focus on shortest or fastest paths alone seems rather far from the situation in many real systems, especially since the path durations are continuous and there may be paths that are only differentially slower.

Another class of centrality measures takes its starting point in the assumption that something diffuses randomly around the network, instead of traveling from source to target along the shortest paths, as for closeness and betweenness. A central vertex in such a setup is a vertex that often is occupied by that something~\cite{hill_braha}. For static graphs, this approach yields matrix-based centrality measures like the \emph{eigenvector centrality}, \emph{Katz centrality} and \emph{PageRank}~\cite{easley_kleinberg,newman_2010}. Since this review focuses on methods that do not aggregate the temporal dimension, any generalization of these measures worth mentioning would have to use three-dimensional tensors representing the temporal network. Instead of working through tensor algebra, we describe the algorithm of a generalization of the eigenvector centrality.
\begin{enumerate}
\item Start with a centrality value 1 at each vertex.
\item At every contact between vertices $i$ and $j$, let the $C_E$ values after the contact at timestep $t$ be
\begin{equation}\label{eq:flow_i}
C_E^{t+1}(i)=\zeta C_E^t(i)+(1-\zeta)C_E^t(j)
\end{equation}
and
\begin{equation}\label{eq:flow_j}
C_E^{t+1}(j)=\zeta C_E^t(j)+(1-\zeta)C_E^t(i) .
\end{equation}
\end{enumerate}
When the dataset is exhausted, we have a distribution of the centrality that we can take as a generalization of the eigenvector centrality (even though it is not derived from the solution of an eigenvalue equation). As opposed to the shortest-path based measures this one becomes, automatically, time-aggregated. The parameter $\zeta$ sets the rate of centrality transmitted at a contact. If $\zeta$ is large it puts a bigger emphasis on recent contacts. A sensible range of $\zeta$ is the interval $(0,1/2]$, but in practice it is system dependent and one needs to choose it by finding the value for which it ranks of the vertices as well as possible compared to some external data. In the upper limit, the vertices share their collective centrality upon a contact.

Similarly to the eigenvector centrality applied for directed networks  (in an iterative implementation like the one above), the temporal eigenvector centrality algorithm can get stuck at vertices. In other words, the centrality is not updated after the last contact, which means the vertices whose last contact is fairly early in the contact sequence will in the end have a score that has not been updated very recently. A remedy, similar to the PageRank or the Katz centrality, is to add a fixed centrality score to every vertex at every time step. To make the contribution uniform over time, one can normalize the score after every timestep (or, equivalently, increase the added value so that it is a constant fraction of the total centrality score). A version of this approach is discussed in Grindrod \textit{et al.}~\cite{grindrod_etal}.

\subsection{Persistent patterns}

Let us now move beyond temporal network measures related to paths and address temporal patterns and subgraphs. Lahiri and Berger-Wolf~\cite{lahiri_berger_wolf_2007} discuss how to identify subgraphs that are persistent in temporal networks. They define the \emph{support set}  $S(G')$ of a subgraph $G'$ (where $G'\subseteq G_t$ is a subgraph of $G_t$---the graph of all edges active at time $t$ in a temporal graph) as the set of timesteps when $G'\subseteq G_t$. They go on to define a ``frequent'' subgraph that we rather call a \emph{persistent subgraph} as a graph that has a support larger than a certain threshold value. The authors have subsequently developed a method of detecting periodic patterns among subgraphs using the support set concept~\cite{lahiri_berger_wolf_2008}.

\begin{figure}
\includegraphics[width=0.9\linewidth]{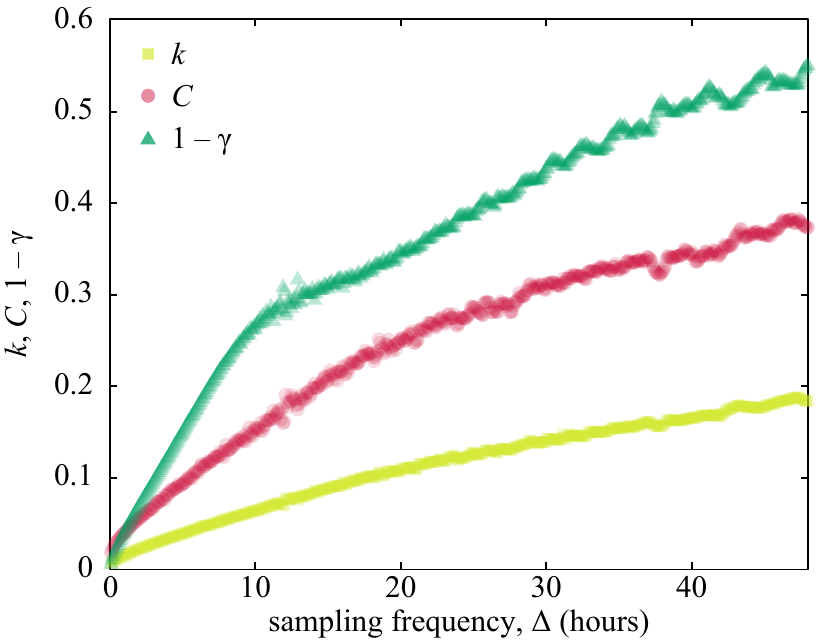}
\caption{Three static network quantities as a function of the sampling time window $\Delta$ (from Ref.~\cite{clauset_eagle}). $k$ is the average degree (twice the number of edges per vertex); $C$ is the clustering coefficient (the number of triangles normalized to the unit interval given the number of connected triples of vertices, see Ref.~\cite{newman_2010}); $\gamma$ is the adjacency correlation coefficient of Eq.~\ref{eq:adjacency_corrrelation_function}.}
\label{fig:clauset_eagle}
\end{figure}

In another study of persistent patterns in a temporal proximity network, Clauset and Eagle~\cite{clauset_eagle} investigated time slices of an interval graph representation as a function of the duration of the time window $\Delta$. They found that the static network structure of the time slices depends much on $\Delta$---not surprising perhaps since the average degree grows with $\Delta$---but some quantities shows abrupt changes of their $\Delta$-scaling indicating characteristic time scales of the behavior of the sampled people. This is illustrated in Fig.~\ref{fig:clauset_eagle}. One of the curves in this figure corresponds to a nifty measure of the similarity in the connection pattern of a vertex $i$ by the \emph{adjacency correlation function} (also called \emph{temporal-correlation coefficient} in \cite{tang_etal_2010})

\begin{equation}\label{eq:adjacency_corrrelation_function}
\gamma_i(t)=\frac{\sum_{j\in\phi(i,t)}a(i,j,t)\,a(i,j,t+1)}{\sqrt{\sum_{j\in\phi(i,t)}a(i,j,t)}\sqrt{\sum_{j\in\phi(i,t)}a(i,j,t+1)}}
\end{equation}
where $t$ represents the entire time window and $\phi(i,t)$ is the set of indices $j$ such that $a(i,j,t)$ or $a(i,j,t+1)$ is one. Fig.~\ref{fig:clauset_eagle} displays the degree $k$, the clustering coefficient $C$ and $\gamma_i(t)$ averaged over $i$ and $t$ as a function of $\Delta$. 
\subsection{Motifs}

Another approach for detecting significant patterns in networks is related to \emph{motifs}. A network motif~\cite{alon} is an equivalence class of subgraphs that is overrepresented in terms of its cardinality with respect to some null model in a network, i.e.\ a larger number of such subgraphs can be found than in a randomized reference system \footnote{It should be noted that the word \emph{motif} is often used in other meanings than an overrepresented class of subgraphs---equivalence classes such as feedforward triangles may be called motifs whether they are overrepresented or not. In addition, the word motif is at times used to denote the subgraphs that constitute a motif}. Usually, the configuration model that conserves the degree sequence but otherwise randomizes the network is taken as the reference system. The over- or underrepresentation of certain subgraphs can be related to the function of the system, especially in directed networks where the subgraphs forming motifs can be associated with e.g.\ information processing tasks. 

There are several ways to extend this concept to temporal networks. The easiest is to look at snapshots of the network taken at different points in time, or alternatively aggregated edges over a period of time, and count the different subgraphs in these snapshots. Braha and Bar-Yam~\cite{braha_bar_yam} does exactly this for an email data set, where one snapshot network represents contacts aggregated over a day. They look at motifs of four vertices and compare them to an edge rewiring null model. The main conclusion from Braha and Bar-Yam's study is that the denser motifs are overrepresented. This approach, however, projects out the information from the order of events, as the motifs are no longer temporal networks themselves. In Ref.~\cite{bajardi2011}, Bajardi \textit{et al.}\ apply an even simpler definition for their dynamical motifs that are in essence time-respecting paths constructed from events belonging to adjacent snapshot time windows. Interestingly, when the time-reversal null model is applied on data on cattle movements (see Sect.~\ref{sec:null_models}), the number of such paths is observed to be smaller than for the original data, indicating the presence of an "arrow of time" in the system.

The communication motifs addressed in Zhao \textit{et al.}~\cite{zhao_etal} are also defined on the basis of static subgraphs, although temporal information is used for defining the subgraphs of interest. Here, the authors take an approach where communication events in mobile telephone call records or Facebook wall postings are first linked to form a communication graph if they both share vertices and  succeed one another within a time limit $\Delta t$, similarly to the path waiting-time cutoff. The difference to the above definitions is the absence of a snapshot window; temporal adjacency of events is based on the time difference between two events sharing a vertex, instead of aggregating the entire system. Communication motifs are then constructed on the basis of graph equivalence of (static) subgraphs found in such networks. Zhao \textit{et al.}\ observe high frequencies of chain, star, and ``ping-pong'' subgraphs that can hardly be found in reference networks where the event times have been randomly reshuffled (the Randomly Permuted Times null model, see Sect.~\ref{sec:null_models}). Such motifs are seen to be stable over time. Zhao \textit{et al.}\ also define ``maximum flow'' motifs, where call durations are used as a filtering criterion.

Chechnik \textit{et al.}~\cite{chechik_etal} define temporal motifs for gene-regulatory networks controlling metabolic pathways. In this case, the primary temporal unit is the genes, or vertices, not the edges. Their method proceeds by coding the time evolution of the expression level of a gene into a sequence of states and comparing the relative timing of the transitions between theses states for different connected genes. Furthermore Chechnik \textit{et al.}\ use randomization techniques similar to those in Sect.~\ref{sec:randomization} to infer statistically significant ``activity motifs'' as they call them.

\begin{figure}
\includegraphics[width=0.8\linewidth]{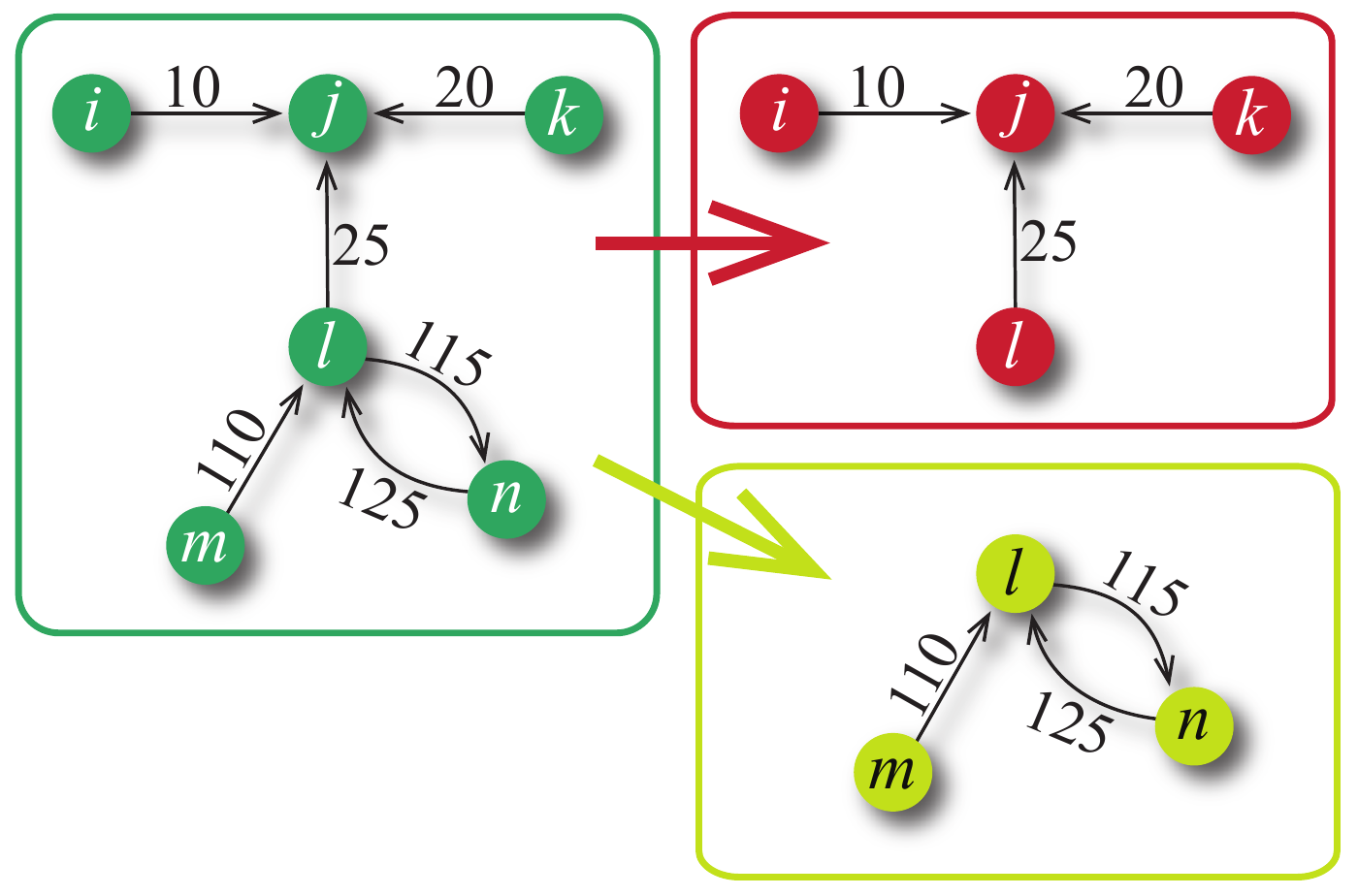}
\caption{A contact sequence (the numbers on the edges denote the times of contacts) involving five vertices (left), and the two maximal $\Delta t$-connected temporal subgraphs found within this sequence when $\Delta t=15$. After Ref.~\cite{Kovanen2011}.}
\label{fig:motif_example}
\end{figure}

In Ref.~\cite{Kovanen2011}, Kovanen \textit{et al.}~define temporal motifs such that the equivalence classes are based on temporal subgraphs, i.e.\ the order of contacts constituting the subgraph/motif matters. The requirements for the temporal subgraphs are based on the following definitions: 
two contact events $e_i$ and $e_j$ are adjacent if they share a vertex, and they are \emph{$\Delta t$-adjacent} if additionally their time difference is no larger than $\Delta t$.
Further, two events $e_i$ and $e_k$ are \emph{$\Delta t$-connected}, if there exists a sequence of $\Delta t$-adjacent events joining $e_i$ and $e_k$. Note that this sequence does not have to be a time-respecting path.  
A temporal subgraph may now be defined as a set of events where all pairs are $\Delta t$-connected. Requiring that no events are skipped at nodes within the subgraph yields \emph{valid} temporal subgraphs. Temporal motifs are now defined as classes of isomorphic valid subgraphs, where the isomorphism is taken to include the similarity of the \emph{temporal order of events}, but not their exact timings. Furthermore, for every event $e_i$ there is a unique maximal $\Delta t$-connected temporal subgraph that includes $e_i$ and the largest possible set of events that are still pairwise $\Delta t$-connected. Motifs based only on maximal temporal subgraphs are called maximal motifs.

For the algorithmic solution, such temporal subgraphs may nevertheless be mapped into static directed graphs, whose isomorphism can then be used for computing subgraph counts for the temporal motif equivalence classes. One can further merge equivalence classes by considering (potential) information flow within the event sequence of a temporal subgraph---in some cases, there are some events whose detailed order does not matter. When applying these definitions and a corresponding algorithm to large-scale mobile call data, Kovanen \textit{et al.}~found out that temporal motifs where the event sequences may have a causal explanation are more frequent than sequences where the event order appears random.

Regarding temporal motifs, there are issues that still remain unresolved and would warrant further consideration. A very important issue is that of reference or null models. For static networks, motif analysis is concerned with over- or underrepresentation of subgraphs when compared with a reference ensemble where structural correlations have been removed. However, what the reference ensemble should be for temporal motifs is far from obvious, as it is evident that for temporal graphs where the events are fairly sparse, a much smaller number of temporal motifs can be found in a reference ensemble where the event times have been randomly reshuffled. Thus any larger temporal subgraphs where the waiting times between events are short are almost always overrepresented, and not much information is gained from applying the null model. Another issue that would deserve further attention is the recurrence of temporal subgraphs---either exactly the same or to some extent similar sequence of contact events may take place between the same set of vertices at multiple points in time. Understanding such recurrent mesoscale patterns might yield a lot of insight e.g.\ on the function of social communication networks, or functional brain networks. Furthermore, for some systems, studying the time dependence of the occurrence counts of temporal subgraphs might also yield useful information.

\subsection{Measuring inter-contact times and burstiness}

In addition to measures characterizing the temporal networks and patterns in them, it is often also useful to have a closer look at the smallest building blocks of such networks -- the vertices and edges, and the associated sequences of contacts. In a typical temporal network, there are multiple contacts between connected vertices, and correlations in the timings of such contacts play an important role \emph{e.g.}~regarding durations of time-respecting paths and latencies between vertices. Especially, for temporal networks of human communication, it has been discovered that such timings are often \emph{bursty} and deviate from the more uniform times expected from a memoryless, random Poisson process~\cite{eckmann_etal,johansen,barabasi,vazquez_etal,oliveira_barabasi,iribarren_moro,candia_etal,wu_etal,karsai_etal,miritello_etal,malmgren_etal_2009}. Overall, the burstiness manifests itself in broader-than-expected distributions of inter-contact times, $P(\tau)$, defined either for nodes or their individual edges. However, because of inhomogeneities and broad distributions of node degree, numbers of contacts per node, and numbers of contacts per edge, directly interpreting the shape of $P(\tau)$ can be difficult as it reflects combined effects of burstiness and such inhomogeneities. Because of this, displaying \emph{scaled} inter-contact time distributions has become the norm~\cite{candia_etal, miritello_etal,karsai_etal}. Here, nodes (edges) are binned according to their numbers of contacts, and the inter-contact time distributions are scaled by the average inter-contact time in each bin, $P(\tau/\tau^\ast)$. When compared against similar distributions for uniformly random contact times, broad tails are typically observed. 

Another alternative is to directly derive a measure of burstiness for any sequence of inter-contact times. Here, the statistical properties of inter-contact times arising from a Poisson process form a natural point for comparison. Goh and Barab\'asi~\cite{Goh08} use as their starting point the \emph{coefficient of varition}, defined as the ration of the standard deviation of the inter-contact times to their mean, $\sigma_\tau/m_\tau$. For a Poissonian contact sequence, $\sigma_\tau/m_\tau=1$. Using this quantity, the \emph{burstiness} of a sequence is then defined as
\begin{equation}
B=\frac{\left(\sigma_\tau/m_\tau-1\right)}{\left(\sigma_\tau/m_\tau+1\right)}=\frac{\left(\sigma_\tau-m_\tau\right)}{\left(\sigma_\tau+m_\tau\right)}.
\end{equation}
For finite contact sequences, the variance is always finite, and $B\in(-1,1)$, such that $B=1$ indicates a most bursty sequence, $B=0$ a sequence with Poissonian inter-contact times, and $B=-1$ a completely periodic sequence.

\subsection{Entropies and other information-theoretic measures}

Information theory is the branch of science exploring the limits of storing, communicating and compressing data. It has already found applications in the network-clustering literature (e.g. Ref.~\cite{rosvall_bergstrom}) and could well be extended to temporal networks. One such attempt is Timo,  Blackmore and Hanlen's entropy-based method to account for temporal uncertainties in communication networks~\cite{timo_etal}. While this is not a deterministic temporal-network approach, as in most of this review, it points to an interesting direction for future research. Similar ideas have been developed in the analysis of ecosystems. Ulanowicz define a measure ``ascendency'' to quantify how well a system process its environmental input~\cite{ulanovicz,pahlwostl}. The measure, derived from information theory, can be adapted to ecological interactions changing in time.

\section{Representing temporal data as a static graph}
\label{sec:static}

The literature on static graphs is many times larger than that on temporal graphs, for a natural reason: it is usually much easier to analyze static graphs, especially analytically. One approach to analyzing temporal graphs is thus to derive static graphs that capture both temporal and topological properties of the system. The most straightforward way is to accumulate the contacts over some time to form edges. This can be an informative approach to studying the time evolution of the static structure of an evolving network~\cite{carley}. Typically, authors have either investigated a network-topological quantity as a function of time when every contact between a pair of vertices that has not been in contact before adds an edge to the graph (e.g.\ Ref.~\cite{holme_2003}), or they have divided the time into segments and studied the structure of contacts accumulated over those segments~\cite{rosvall_bergstrom}. As mentioned above, this trivial way of projecting out the temporal dimension can discard information and this review focuses on methods that attempt to capture it. Nevertheless, it is useful whenever the topological aspects are more important than the temporal. It may also be possible to combine these aspects: Miritello \textit{et al.}~\cite{miritello_etal} have proposed a way of mapping the dynamic SIR model to a static edge percolation model, where the edge weights of a network whose structure equals that of the aggregated graph are defined in a way that takes into account the temporal correlations and inhomogeneities of edges (see Section~\ref{sec:compartmental}). 
There are other ways of encoding the temporal network structure into a static graph different than the aggregated graph that incorporate more temporal information than a plain projection of a contact sequence or interval graph into the graph dimension.

\begin{figure}
\includegraphics[width=0.9\linewidth]{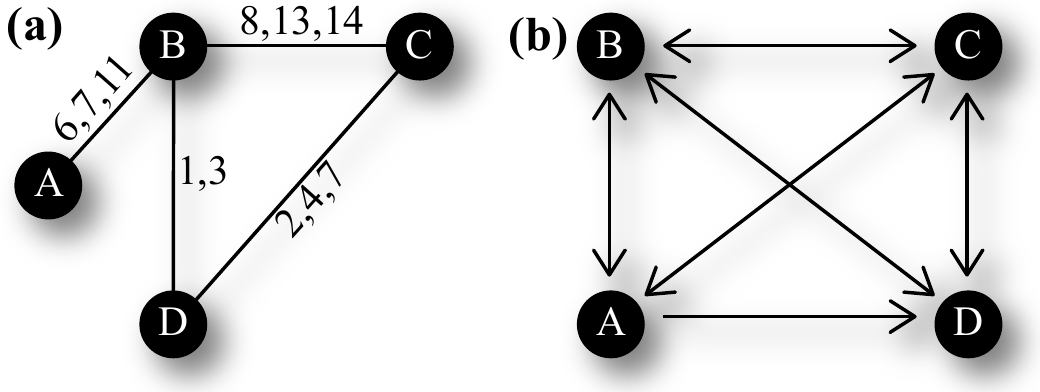}
\caption{Reachability graphs. Panel (a) shows a contact sequence (same as in Fig.~\ref{fig:reachability}) and (b) shows its reachability graph.}
\label{fig:reachability_graph}
\end{figure}

\subsection{Reachability graphs}

An alternative graph representation that might be useful at least for very sparsely connected contact structures is reachability graphs, or ``path graphs''~\cite{moody}, or ``associated influence digraph''~\cite{cheng_etal}. In such a case one puts a directed edge from vertices A to B if there is a time-respecting path from A to B (see Fig.~\ref{fig:reachability_graph}). Such a graph, thus, shows which vertices can possibly affect which others. The average degree $k$ of a reachability graph is thus the average worst-case outbreak size minus one. In other words, for any contact structure that supports a pandemic, the reachability graph will be dense ($k \sim N$). This is a drawback for reachability graphs since most methods to analyze networks are developed for sparse graphs. Nevertheless, one can imagine very sparse connection structures that can benefit from such a representation.

One work using reachability graphs is Bearman \textit{et al.}~\cite{bearman_etal} that studies dating between high-school students. Their data fits the interval-graph framework (Fig.~\ref{fig:types}b) where contacts can happen at any time during the intervals, and the only thing one knows is that the contact happens at least once. In such a case the reachability graph is not unique and the authors plot a maximal and minimal reachability graph. Going in the opposite direction, one can prove that given a reachability graph, one can always construct a temporal graph that has the given reachability structure~\cite{cheng_etal}.

\subsection{Line graphs}

Although it is a bit outside the scope of this review, we mention that line graphs have been used to study disease spreading in temporal networks represented by aggregated time slices~\cite{liljeros_edling_amaral}. A line graph of a simple static graph $G$ is a graph whose vertices are the edges of $G$ that are connected if they share a vertex in $G$. Sometimes the line graph is called ``interchange graph'' or ``dual graph'' (the latter being a bit of a misnomer since the dual of the dual of $G$ is not $G$). The point of studying line graphs in epidemiology is that they are closely related to the structure of concurrent partnerships---e.g.\ the number of edges of the line graph is the number of concurrent partnerships in the original graph. 

\subsection{Transmission graphs}

\begin{figure*}
\includegraphics[width=0.75\linewidth]{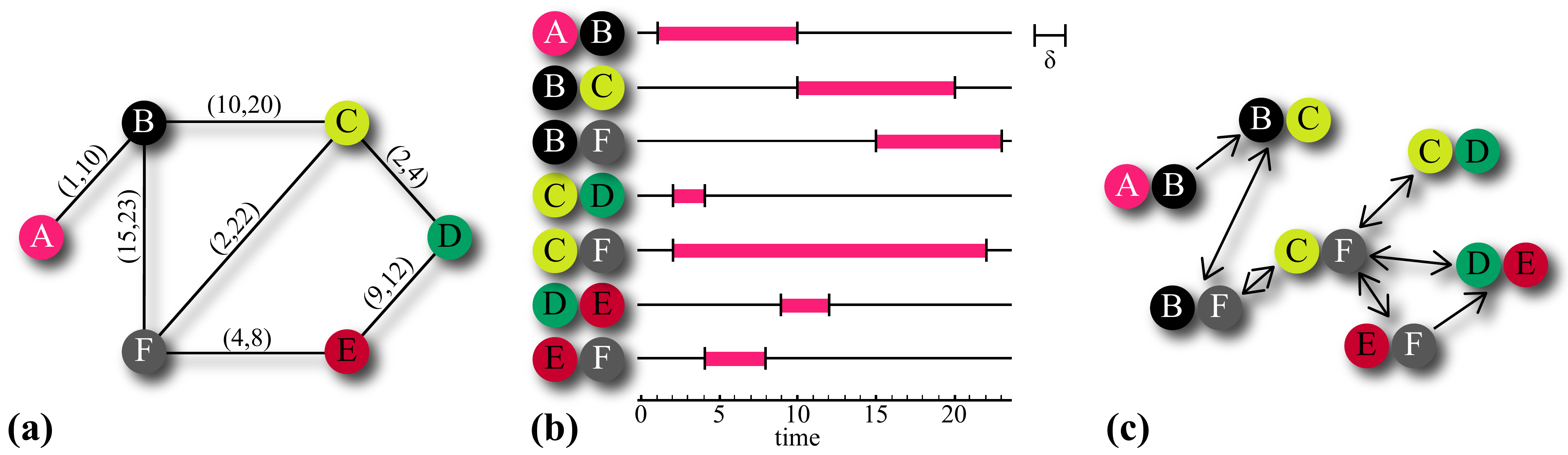}
\caption{Transmission graphs. Panel (a) shows an interval graph representation of a temporal network (where each edge has only one interval, as required by the definition by Riolo \textit{et al.}~\cite{riolo_etal}). Panel (b) gives slightly reduced picture of the system, where a line corresponds to an edge in (a) and the active interval is indicated. The derived transmission graph is illustrated in (c).}
\label{fig:transmission_graph}
\end{figure*}

Riolo \textit{et al.}~\cite{riolo_etal} present a version of line graphs adding some more temporal information. What they call a \emph{transmission graph} thus goes beyond modeling the contact structure and also incorporates the disease dynamics via a parameter $\delta$ that quantifies the combined incubation time and duration of a disease. Their starting point is a graph where an edge $e$ is active over an interval $[t_{\textrm{start}}(e),t_{\textrm{stop}}(e)]$. There is a directed edge from $e$ to $e'$ in the transmission graph if $e$ and $e'$ share a vertex, $t_{\mathrm{start}}(e) < t_{\mathrm{start}}(e') + \delta$ and $t_{\mathrm{start}}(e') < t_{\mathrm{start}}(e)$. See the illustration in Fig.~\ref{fig:transmission_graph}. The idea with the extra $\delta$ is to identify the last possible time a member of an edge can transmit the disease. The advantage over a pure line graph is that transmission graphs encode the directionality arising from the order of non-concurrent relationships. However, in common to line graphs of concurrent relationships, the transmission graphs cannot handle edges where one vertex manages to not catch the disease. Furthermore, paths in transmission graphs do not have to be time respecting. Riolo \textit{et al.}~\cite{riolo_etal} define some different versions of their transmission graphs cutting the contact sequence in various ways, but essentially the idea is outlined above.

\section{Models of temporal networks}

In general, models of networks can serve many purposes. They may explain the emergence of salient network characteristics, or serve to produce synthetic networks with desired, tunable characteristics that can then be used in computational experiments of dynamic processes. Another class of models is that
of randomized reference networks, where empirical networks are taken as inputs and randomization procedures are used to remove some of their characteristic correlations. For temporal networks, the number of models proposed in the literature is still fairly limited. Below, we will first discuss some models proposed for temporal social networks and epidemiological contact networks, and then move on to randomized reference models. 

Note that the word "model" has a slightly different meaning when used in the context of statistical inference -- here, techniques of inference and machine learning are used to extract statistical models, typically from limited amounts of data. Methods for statistical inference of temporal networks are still rare, although
this problem is an important one for systems biology, e.g.~in the context of temporal networks of gene regulation, where Ref.~\cite{lebre_etal} use a Markov Chain Monte Carlo approach to infer the parameters of a so-called Bayesian Network representation of gene regulatory networks~\cite{lebre}. This approach could probably have a wider use for systems also outside of biology.

\subsection{Models for temporal social networks}

\subsubsection{Temporal exponential random graphs}

The method of exponential random graphs (see, e.g.~\cite{robins_etal}) is commonly used by social scientists. The parameters of such graphs, inferred from empirical data, contain information on the importance and frequency of chosen topological elements and subgraphs, such as triangles and stars. A similar modeling framework for temporal exponential random graphs has been put forward in e.g.\ Refs.~\cite{hanneke_xing,guo_etal,kolar_etal}. The basic problem in this literature is that given an observed time evolution of a set of states of the vertices, representing some dynamical system on the graph, one should determine the parameters of a time-varying exponential graph model that essentially gives the probability of a contact to happen between a pair of vertices at a given time. Just like its static counterpart---exponential random graphs---has a connection to the Ising model, the temporal exponential random graphs can also be boiled down to finding a partition function, in this case a time-varying one. When this is done, the temporal exponential random graph model can be used both as a reference model for measuring biases in quantities of topological and temporal structure, as well as a generative model for tuning the structure of contact sequences for simulations of dynamical systems on top of these.

\subsubsection{Models of social group dynamics}

Zhao, Stehl\'e, Bianconi and Barrat have presented a framework for modeling social networks~\cite{stehle_etal,zhao_etal_2011} where edges represent ephemeral social ties, such as being in face-to-face contact. Their approach is based on master (or ``rate'') equations that represent the expected change in the number of people in a group of a certain size, and can capture observations like ``the longer an agent interacts with a group, the less it is likely to leave the group; the more the agent is isolated the less likely it is to interact with a group.'' In essence, this framework deals with interval graphs rather than contact sequences, much like the neighborhood exchange model discussed below. 

In contrary, the model by Jo \emph{et al.}~\cite{hangPlosOne2011} explicitly accounts for the contact timings, combining their short time scale with the longer time scale of network evolution. Jo \emph{et al.} propose a social network model that combines features of earlier social network models (focal and cyclic closure, tie strength reinforcement \cite{Kumpula07}) with triad interactions and social interaction task execution from a priority queue, along the lines of Ref.~\cite{barabasi}. This model gives rise to networks with communities of strong ties connected by weak links; the contacts mediated by the links are bursty.

\subsection{Contact network models}


As a simple way of extending static graphs to involve a turnover of neighbors is was proposed by Volz and Meyers~\cite{volz_meyers}. This model is rather similar to the RE randomization procedure discussed below, but with the purpose of mimicking the change of partnerships to generate contact structures for disease simulations. It works by selecting two edges with some probability every time step and swapping them as in step~3 of the RE procedure. As the model is based on rewiring an existing network, the topology of the accumulated network and the contact patterns across edges have to be determined using some other models.


In order to generate synthetic sexual contact networks, Kretzschmar \textit{et al.}~\cite{kretzschmar_etal} have proposed a model that generates interval graphs and allows tuning the assortativity of the resulting networks.. The model rules are as follows:
\begin{enumerate}
\item
\begin{enumerate}	
\item \label{step:form_partnership} A new partnership is formed with probability $\rho$. The individuals participating in the pair are chosen according to the mixing function $\phi$:
\begin{enumerate}
\item draw two random individuals $i$ and $j$;
\item decide whether they form a pair according to $\phi(i,j)$;
\item if yes, done; else go to $i$.
\end{enumerate}
\item Repeat step~\ref{step:form_partnership} $N/2-P$ times.
\end{enumerate}
\item In every pair consisting of a susceptible and an infected, the disease is transmitted with probability $\eta$.
\item Every pair splits up with probability $\sigma$.
\end{enumerate}
The mixing function is introduced to be able to tune the mixing by degree~\cite{newman_2010}. It could be
\begin{equation} \label{eq:assortative}
\phi(i,j)=1-\xi+\xi\frac{k_ik_j}{k_{\mathrm{max}}^2}
\end{equation}
for assortative mixing (where there is a tendency for high-degree vertices to connect to other high-degree vertices and low-degree vertices to low-degree vertices), or
\begin{equation} \label{eq:disassortative}
\phi(i,j)=1-\xi+\xi\frac{(k_i-k_j)^2}{k_{\mathrm{max}}^2}
\end{equation}
to create disassortative mixing where high-degree vertices tend to connect to low-degree vertices. The parameter $\xi$ sets the strength of the assortativity or disassortativity. $k_{\mathrm{max}}$ is an upper limit of the degree---this was before the finding that sexual networks have power-law degree distributions (Liljeros \textit{et al.}~\cite{liljeros_etal_2001}). A modern approach would probably be to draw degrees from a distribution and then connect them with a function like $\phi$. Except controlling for assortativity, one can also choose $\phi$ to create serial monogamous relationships by letting $\phi=1$ if $k_i=k_j=0$, otherwise $\phi=0$. In a paper on measuring concurrency (the temporal overlap of partnerships) Morris and Kretschzmar~\cite{morris_kretzschmar} used the parameter values:  $N=2000$,  $\rho = 0.01 / \mbox{day}$,  $\sigma = 0.001 / \mbox{day}$ and $\eta = 0.1 / \mbox{day}$. We note that, as a model for sexual networks this model fails to capture that overall sexual activity decreases with the number of partners~\cite{nordvik_liljeros}. 

A fair amount of subsequent works has followed ideas similar to the stochastic pair-formation model. We will not review all of them but mention the ``dynamic random graph models with memory'' of Turova~\cite{turova}, which effectively models the same type of time-evolving graphs as above, but use a framework more tractable for analytic calculations. Other recent modeling evolving interval graphs but including more structure are presented in Refs.~\cite{snijders_koskinen_etal,snijders_vandebunt_etal}.

\subsection{Randomized reference models\label{sec:randomization}}
\label{sec:null_models}

\begin{figure*}
\includegraphics[width=0.7\linewidth]{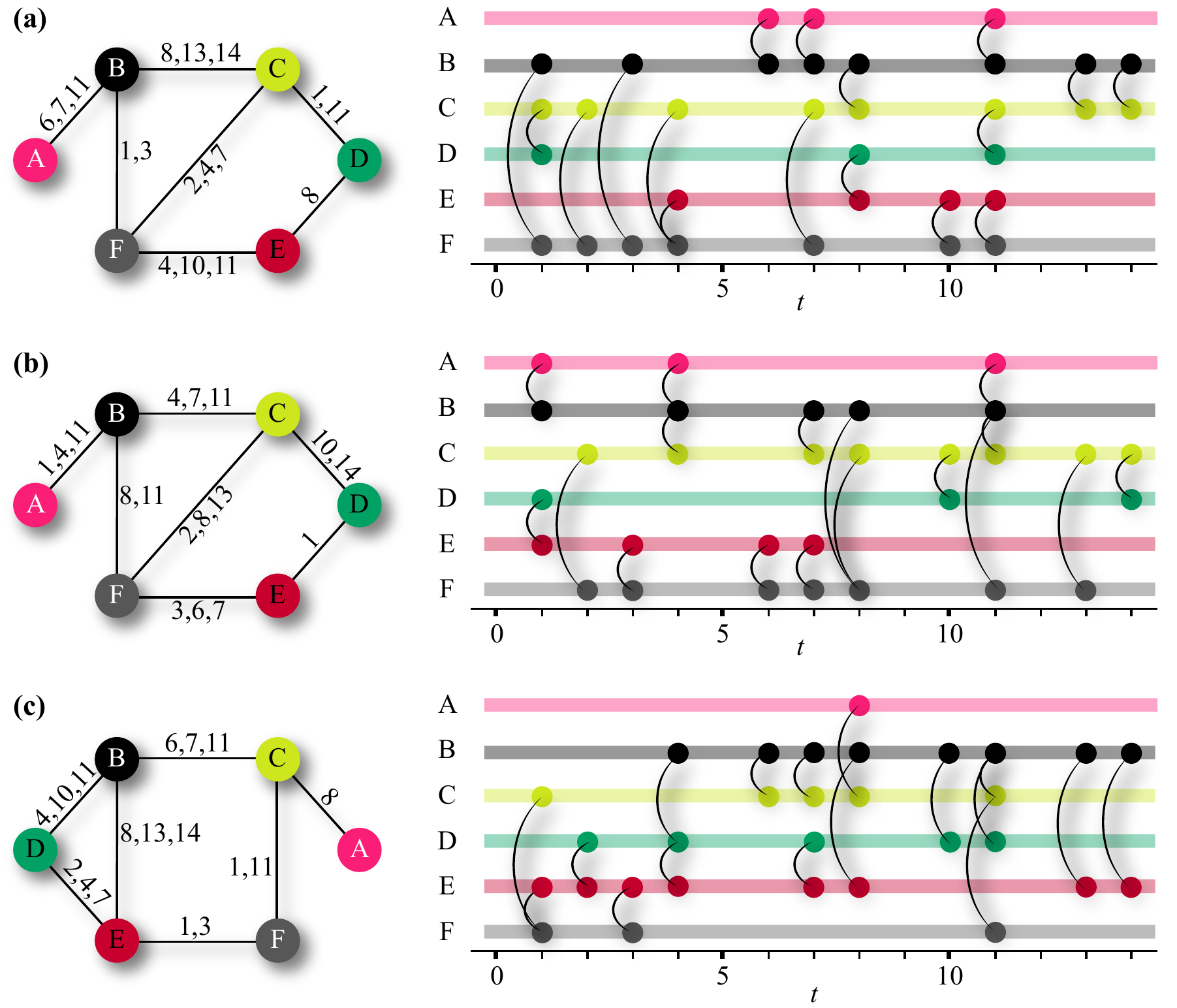}
\caption{Illustration of two types of randomization null-models for contact sequences. (a) shows a contact sequence (the same as in Fig.~\ref{fig:reachability}). In (b) it is randomized by the Randomly Permuted times procedure such that contacts happen the same number of time per edge, and the aggregated network topology is the same. In (c) the contact sequence in (a) is randomized by the Randomized edges (RE) procedure. With RE, the time sequence of the contacts along an edge is conserved, and so is the degree sequence of the original network, but all other structure of the topology is destroyed. (The latter statement is perhaps not so well illustrated by this figure as there are not so many graphs with the degree sequence of the original, aggregate graph.)}
\label{fig:randomization}
\end{figure*}

For static networks, a common way of assessing the importance, unexpectedness, or over/underrepresentation of topological features of empirical networks is to compare the features against some reference model where the network is randomized. The most widely applied reference model is the configuration model, where the links of the original network are randomly rewired pairwise. This reference model preserves the original degree sequence but yields otherwise maximally random networks. Then, one can assess the significance of quantitative topological characteristics of the empirical graph by either a direct comparison to averages in the randomized reference ensemble, by computing $Z$-scores for the characteristics with respect to their distributions in the reference ensemble given by the reference model, or by measuring the extent to which the dynamics of some processes differ when run on the empirical networks and the reference ensemble. 

For temporal graphs, a similar approach can be applied: in this case, the original event sequences are randomized or randomly reshuffled in a chosen fashion to remove time-domain structure and correlations. However, there are several kinds of possible temporal correlations and several time scales where the correlations are important, and thus no single, general-purpose null model  can be designed (Òthe temporal configuration modelÓ). Rather, by designing appropriate null models, one may Òswitch offÓ selected types of correlations in order to understand their contribution to the observed time-domain characteristics of the empirical temporal network. Such temporal null models have also been applied for studying the effects of various kinds of correlations on dynamical processes (such as spreading) on temporal graphs. A typical use for such models in studies of dynamical processes would be to essentially apply all of them, and by monitoring how the dynamics of the process depends on the reference models, to pinpoint the role of different temporal and topological correlations on the process -- if removing a certain type of correlations changes the dynamics a lot, then obviously those play an important role for the dynamics.

Below, we review temporal null models introduced in the literature, essentially following Holme~\cite{holme_2005} and Karsai \textit{et al.}~\cite{karsai_etal}. For this section, we assume that the temporal network is a contact sequence. Some of the methods work for interval graphs too; others can be modified to interval graphs quite straightforwardly. In the end of the section, we summarize and provide some guidelines for choosing reference models.


\subsubsection{Randomized edges (RE)}

This method is similar to the configuration model for static graphs mentioned above, with the additional ingredient that contact sequences of edges follow the edges when these are rewired. Algorithmically, the method is defined as follows:
\begin{enumerate}
\item \label{step:loop} Go over all edges sequentially.
\item \label{step:pick} For every edge $(i,j)$, pick another edge $(i',j')$.
\item \label{step:replace} With a probability $1/2$ replace $(i,j)$ and $(i',j')$ by $(i,j')$ and $(i',j)$, otherwise replace them by $(i,i')$ and $(j,j')$.
\item If the move in step~\ref{step:replace} created a self-edge or multiple edge, then undo it and start over from step~\ref{step:pick}.
\end{enumerate}
The times of contact over an edge are kept constant. Note that the two alternatives in step~\ref{step:replace} where one is randomly selected are needed to remove spurious correlations if the data structure that is used returns the vertices of an edge in a specific order; otherwise one would keep the number of times a vertex appears in the first argument conserved, which in practice can give quite big differences for empirical graphs, whether the graph is small or not. To speed up the process, one can skip edges that already have been rewired in step~\ref{step:loop} (by being selected in a previous step~\ref{step:pick}). On the other hand, this procedure is linear in $M$ and rarely a computational bottleneck.

This null model can be used to study the effect of the network topology, that is, the wiring diagram of the original network. The model also assumes that it is the edges rather than the vertices that govern the times of contacts---after the randomization procedure, both the numbers and timings of contacts for each vertex will have changed; however their degrees in the aggregated network are retained. As the contact sequences follow their edges when rewiring, all temporal correlations and inhomogeneities associated with individual edges, such as burstiness and the distribution of inter-contact times of edges, are retained, as is the overall event rate at every point in time. The RE procedure is illustrated in Fig.~\ref{fig:randomization}.

\subsubsection{Randomly permuted times (RP)}

As a temporal counterpart to the configuration model, one can permute the contact times randomly while keeping the network structure and the numbers of contacts between all pairs of vertices fixed. Technically, this is much simpler than applying the edge rewiring scheme discussed above and it only requires randomly exchanging the time stamps of all contacts, or just randomly reshuffling the order of time stamps in an array or vector. No checks similar to step 4 of the RE rule need to be performed for contact sequences; for interval graphs, one should check for non-overlap. As this null model retains all network structure and the number of contacts for each edge, its application can be used to study the effects of the detailed order of events, including burstiness, inter-contact time distributions of contact sequences on vertices and edges, and correlations and triggering effects between contacts on adjacent edges. The model also retains the overall rate of events in the network at every point in time, such as daily or weekly patterns in communication networks. An illustration of RP can be found in Fig.~\ref{fig:randomization}.

\subsubsection{Randomized edges with randomly permuted times (RE + RP)}

This null model works as follows: first, the network structure is randomized using the RE procedure. Then, the time stamps of all contacts are reshuffled with the RP scheme. Thus the outcome is a temporal graph, where all structural correlations and all temporal correlations with the exception of the overall rate of contacts (such as daily/weekly pattern) have been destroyed.

\subsubsection{Random times (RT)}

The RP ensemble conserves the set of times of the original contact sequence. Hence, although it destroys burstiness of events on individual vertices and edges as well as correlations between events such as triggered chains, the aggregated rate of events in the network is unchanged and will still follow the typical circadian and weekly patterns of human activity~\cite{leland_wilson,holme_2003,malmgren_etal_2008,malmgren_etal_2009,jo_etal,zhou_etal}. So far, results indicate~\cite{karsai_etal} that such overall modulation of the event rate does not appear to matter at least for the simple SI spreading model, whose dynamics is dominated by edge burstiness (see Jo \textit{et al.}~\cite{jo_etal} for a method that removes the contribution of such circadian patterns). However, there are cases where such patterns might play a role, such as more complicated spreading models (e.g.\ SIS or SIR). The random times (RT) null model destroys such patterns: in this approach, each contact on each edge is assigned a random time within the observation time window of the original data, and thus the network structure and the total number of events on each edge are conserved. When comparing the results of any dynamics taking place on the RP and RT models, the difference should then be due to the network-level temporal patterns of the empirical data. Note that another alternative for uniformly random contact times is generating them from any chosen distribution or process, such as the Poisson process~\cite{karsai_etal}, with parameters set up so that the numbers of contact per each edge are on average conserved. 

The above approach assumes the degree distribution is external, as it is only investigating ensembles where the topology and the degree of each vertex is conserved. Another option would be to run the dynamics on Poisson random graphs (Newman~\cite{newman_2010}) with the same number of vertices and edges as the original graph and time stamps distributed as the original data, permuted like RP and random like RT. 

\subsubsection{Randomized contacts (RC)}

Here, one keeps the graph topology fixed but redistributes the contacts randomly among the edges. After this randomization, the number of contacts per edge follows the binomial distribution rather than some broad, right-skewed distribution as is typical for empirical data. This randomization can be utilized in testing the effect of the distribution of the number of contacts per edge in combination with the order of events. Hypothetically, one would like to test the effect of the distribution of the number of contacts alone, keeping the structure of the temporal order of the real data. For example, a vertex that is active primarily in the early stage of the data would be so in the randomized data. This would need a more elaborate approach.

\subsubsection{Equal-weight edge randomization (EWER)}

Karsai \textit{et al.}~\cite{karsai_etal} use a special null model that is designed for removing timing correlations between the contact sequences of adjacent edges, while retaining temporal characteristics associated with edges, including the distribution of inter-contact times on individual edges. Whole contact sequences associated with edges, i.e.\ all contacts and their time stamps, are randomly exchanged between edges that have the same number of contacts. Thus single-edge patterns, such as burstiness, are retained, together with all properties retained by the RP model (number of contacts on each edge, network-level event frequency patterns, topological structure). This null model requires a large enough system so that there are enough edges with the same number of events.

\subsubsection{Edge randomization (ER)}

This null model is similar to the EWER model with the exception that the sequences can be exchanged between edges that have any numbers of contacts. This corresponds to randomly exchanging the edge weights (measured as numbers of contacts) in the aggregated network spanned by the contact sequence, and thus weight-topology correlations are destroyed in the null model. However, the inter-contact time distributions of contact sequences of edges are not changed---the sequences are just placed elsewhere in the network.

\subsubsection{Time reversal (TR)}

This null model is designed for assessing the frequency and importance of  causal sequences~\cite{bajardi2011} of contacts, where contacts trigger further contacts, and simply involves running the original event sequence backwards in time. If sequences of consecutive contacts would be caused by temporal correlations alone, similar numbers of such sequences should be observed when time runs forwards and backwards; the lack of such chains in the time-reversed null model compared to the original sequence can be attributed to ``the arrow of time''.

\subsubsection{Summary and guidelines}

The different randomized reference models discussed above retain and destroy specific kinds of topological and temporal correlations, and thus \emph{e.g.} in studies of dynamical processes, they allow for pointing out the importance of various correlations: the most important correlations can be pinpointed by comparing the effects of different randomization models on the dynamics. The RE and RP models permute edges and contact times. Their simultaneous application (RE+RP) destroys all correlations except for patterns in the overall contact rate -- this provides a good starting point for the limiting case of uncorrelated temporal networks, and by additionally randomizing contact times (RT), the overall patterns are also removed. When studying the roles of the exact contact timings on edges and the correlations between adjacent edges, comparing the EWER and ER models to the RT model should do the trick, as the static network features are retained except for weight-topology correlations removed by the ER.

%

\section{Spreading dynamics and compartmental models on temporal graphs~\label{sec:compartmental}}

One of the key insights of network theory is that the underlying network structure can strongly affect dynamic processes that are mediated by the edges. This is especially important for spreading processes, where biological or electronic viruses, rumors, or pieces of information are transmitted through edges of physical contact, social ties, or electronic connections. For such processes, the network structure affects the speed of spreading as well as the extent of spread through the network through features such as short path lengths~\cite{watts_strogatz}, the degree distribution~\cite{pastor_santorras_vespignani_2001,barthelemy_etal}, degree correlations~\cite{boguna_etal}, or community structure and correlations between tie strengths and network topology~\cite{onnela_etal,park_etal}.

Many types of spreading processes can be modeled as compartmental models, where each vertex is in one of the characteristic states of the model~\footnote{Note that in some contexts, compartmental models refer to models where for each state, one has one variable counting the size of the population in that state, instead of models that explicitly deal with individuals like in the models discussed here.}, i.e.\ belongs to one class or compartment~\cite{anderson_may,hethcote}. In the simplest SI model (Susceptible--Infective), individuals are initially susceptible, and become infected at some rate $\mu$ when in contact with infective individuals. In static networks, this process will eventually infect all vertices that can be reached from the source, and thus the network structure affects only the speed of spreading. The SIR model (Susceptible--Infective--Removed) shows richer dynamics. In this model, infected individuals recover and become immune to spreading, either with some probability $\beta$ per unit time or after a fixed period of time. The outcome of the SIR dynamics now depends on the interplay between the two rates (infection rate $\mu$ and recovery rate $\beta$) as well as the network structure. The process may die out, only infecting local clusters of vertices, or percolate through the network in an epidemic fashion, such that a substantial fraction of individuals is infected. Beyond these two simple models, more complicated versions have been formulated, such as the SIRS model, where immunity is not permanent but individuals may again become susceptible. 

Applying such models to static networks is equivalent to assuming that all interactions between vertices take place uniformly in time. However, in reality this is usually not the case. The spreading of biological viruses depends on temporal patterns of contact, and the flow of information in social networks is influenced by the social activity patterns and rhythms of individuals. Importantly, there is an increasing body of evidence of very large heterogeneity in the timings of such interactions, from burstiness in patterns of communication via electronic and physical mail~\cite{eckmann_etal,johansen,barabasi,vazquez_etal,oliveira_barabasi,iribarren_moro}, mobile telephone calls and text messages~\cite{candia_etal,wu_etal,karsai_etal,miritello_etal}, instant messaging~\cite{leskovec_horvitz}, and patterns in proximity dynamics measured with RFID sensors~\cite{cattuto_etal,stehle_etal_2011,isella_etal}. Daily and circadian rhythms of human dynamics~\cite{leland_wilson,holme_2003,malmgren_etal_2008,malmgren_etal_2009,jo_etal} may also play a role. Furthermore, there are correlations where interaction events trigger further events, such as reception of an email triggering a forwarding event, or incoming calls causing outgoing calls in mobile call networks~\cite{iribarren_moro,karsai_etal,miritello_etal,pan_saramaki}. An overview of dynamic processes important for infectious disease spreading can be found in Bansal \textit{et al.}~\cite{bansal_etal}.

The above temporal inhomogeneities, together with the fact that spreading processes have to follow the time ordering of events, have important and at times drastic effects on the dynamics of spreading on temporal graphs. In the recent years, such effects have been studied with the help of simulations building on empirical contact or interaction sequences, as well as with analytical tools. Although much remains unknown, some clear conclusions can already be drawn from the existing literature, such as the importance of burstiness in slowing down epidemic-style spreading dynamics.

\subsection{Bursty event dynamics and slow spreading in communication networks}

Human communication dynamics is almost universally bursty, i.e.\ the timings between communication events deviate largely from uniform or Poissonian statistics and can instead be described with heavy-tailed or power-law distributions. Consequently, the dynamics of spreading processes also differs from expectations. This failure of the Poissonian approximation (that is typically assumed to hold for spreading models on static networks) was first addressed by Vazquez \textit{et al.}~\cite{vazquez_etal}, who studied email activity patterns from university and service provider logs. 

Their study was motivated by the fact that the numbers of new infections by computer viruses and worms that spread through emails should decay exponentially, if the Poisson approximation was accurate. However, reports of new infections are typically still published years after the release of anti-viruses. In the Poisson approximation, the probability of one vertex to interact with another within a time interval $dt$ is $dt/\langle\tau\rangle$, where $\langle\tau\rangle$ is the average interevent time. Thus, the time between consecutive contacts is exponentially distributed, $P(\tau)\sim\exp(-t/\langle\tau\rangle)$. However, for the email data studied by Vazquez \textit{et al.}, after correcting for finite observation window, it was seen that $P(\tau)\sim 1/\tau$, followed by an exponential cutoff. 

The Poissonian and power-law inter-event time distributions result in different generation times, that is, times between one vertex becoming infected and transmitting the infection to its neighbor. On the basis of generation times, Vazquez \textit{et al.}~\cite{vazquez_etal} calculated an analytical approximation for the prevalence dynamics of the SI model assuming a treelike structure, and showed together with simulations using empirical email sequences that the late-stage exponential decay of the number of new infections is significantly slower for the broad inter-event time distribution and matches well with data on real computer worms.

Later, Min, Goh and Vazquez~\cite{min_etal} studied the SI model with power-law inter-event time distributions using theoretical arguments that assume a tree-like network structure, as well as with simulated spreading with uncorrelated power-law activity patterns and the priority-queue network model. They concluded that a power-law waiting time distribution leads to a power-law decay in the number of new infections in the long time limit, with an exponent determined through the generation time distribution.

More empirical evidence for slower-than-Poissonian spreading dynamics was provided by Iribarren and Moro~\cite{iribarren_moro, iribarren2011}, who performed a viral marketing experiment with emails, where 31,183 individuals forwarded recommendations. They concluded that the large heterogeneity found in the response times is responsible for the slow dynamics of information at the collective level. In their experiment, subscribers to an online newsletter were rewarded for recommending it via email to their friends. The viral spread of this recommendation email was tracked at every step, providing a detailed view on reception and forwarding events and their temporal correlations. As typical for such cascades, the average number of secondary cases per infected individual $R_0$ was below the tipping point $R_0=1$, in this case $R_0\approx 0.26$, and thus all cascades stopped in a finite number of steps. The cascades were observed to be treelike, with a very low clustering coefficient. It was seen that there is a large variability to the number of forwarded recommendation emails and it was argued that this variability is not directly related to the degree of individuals in the email network. Furthermore, those individuals who became secondary spreaders, \emph{i.e.}\ forwarded the recommendation, typically forwarded all their recommendation emails simultaneously in a single spreading event, and did not remain spreaders for longer. The response times (times between reception and forwarding) were seen to follow a lognormal distribution. Traditional analytic epidemic models were seen to fail in predicting the speed and dynamics of the number of individuals who received and forwarded the message; according to the simple exponential growth equation, most new infections should happen during the early few days, whereas a significant fraction of new infections was observed even at the time scale of months. However, the observed slow dynamics was well captured by the non-Markovian Bellman-Harris branching model~\cite{harris} where the lognormal distribution of response times was used. 

Fernandez-Gracia et al.~\cite{graciaPRE2011} studied the effect of broad inter-event time distributions on the ordering dynamics of the Voter model; similarly to the slowing down of compartmental spreading models, such distributions were observed to give rise to slow ordering dynamics of the model.

\subsection{Burstiness and other temporal and structural inhomogeneities}

Karsai \textit{et al.}~\cite{karsai_etal} provided further insight into the effect of temporal heterogeneities on spreading dynamics by studying the behavior of the SI model. The model was simulated with real mobile telephone call and email contact event sequences, together with reference models that destroy selected correlations (see Section~\ref{sec:randomization}). The largest data source was the call database of a mobile telephone operator, containing about $325$ million time-stamped call records over a period of 120 days. For assessing the importance of different temporal and structural inhomogeneities on the spreading dynamics, several reference models were used (with reference to the Section~\ref{sec:randomization}): RP, EWER, ER, a combination of RE and RP, and RT with independent Poisson processes on each edge. As in the above papers, inter-event times were seen to have a broad distribution; these were studied by calculating the distributions for edges binned by event numbers and rescaling those by the average inter-event time in each bin, similarly to Candia \textit{et al.}~\cite{candia_etal} and Miritello \textit{et al.}~\cite{miritello_etal}. It was also seen that the scaling breaks down for small times, around $20$ seconds, indicating correlations and triggering of events.  When compared to the original event sequence, the times to full prevalence (100\% of vertices infected) were seen to be shorter for all null models. The RP model that destroys burstiness and correlations, except for daily patterns, gave faster spreading than the ER model that destroys weight-topology correlations but retains burstiness. Hence, the slowness of spreading can largely be attributed to the bursty event sequences on individual edges, in addition to the ``weak-edge'' bottleneck, i.e.\ edges between communities having a smaller contact frequency. However, the overall daily pattern where the system-level call frequency is low at night and peaks around noon and early evening was seen not to contribute significantly, when investigated with a Poissonian event-generating (RT) model. In addition, the EWER scheme that destroys triggering, i.e.\ temporal correlations between adjacent edges, was seen to result in slightly slower-than-original spreading for short time scales, and slightly faster-than-original for long time scales.  

Interestingly, in contrast to these observations, Rocha \textit{et al.}~\cite{rocha_etal_2011} found that temporal order and correlations speed up epidemic spreading in their data set~\cite{rocha_etal_2010} of sexual contacts in Internet-mediated prostitution -- the origins of this difference still remain unclear. Furthermore, in contrast to both findings, Stehl{\'e} \emph{et al.}~found that when simulating the SEIR spreading dynamics on a temporal contact network of conference attendees recorded with proximity sensors of the SocioPatterns RFID platform~\cite{cattuto_etal}, the spreading dynamics is well described by a static aggregated network if the heterogeneity of the contact durations is taken into account as edge weights.

Miritello \textit{et al.}~\cite{miritello_etal} also used mobile telephone call records (9 billion time-stamped calls of 20 million users over 11 months) in their studies of simulated information spreading. They applied the SIR model with a homogeneous and deterministic recovery time and variable transmission probability $\lambda$ with the empirical call sequence. The scaled distribution of relay times $\tau_{ij}$, that is, times between user $i$ participating in a call with any other user and user $i$ participating in a call with user $j$, was seen to be heavy-tailed while also displaying a larger number of short relay times than expected from the Poissonian case. Similarly to Ref.~\cite{karsai_etal}, the abundance of short relay times was interpreted as a signature of group conversations, where calls trigger further calls and the activity patterns of adjacent edges are thus correlated.

For SIR dynamics, such group conversations and correlated contact sequences were observed to give rise to larger spreading cascades for small values of $\lambda$ below the percolation point than for the time-shuffled reference model where the distribution of relay times approaches the Poisson distribution. However, for large values of $\lambda$, the opposite behavior was observed. Thus, in the context of information propagation, correlations from group conversations make information spreading more efficient at the local small scales when a more realistic low transmissibility $\lambda$ is used. Miritello \textit{et al.}~\cite{miritello_etal} also proposed a way of mapping the dynamic SIR model to a static edge percolation model, similar to Newman~\cite{newman_2002} or Kenah and Robins~\cite{kenah_robins}. They validated their approach by successfully predicting the percolation threshold for the SIR model on empirical data. This was done approximate the average number of secondary infections by an effective, dynamic transmissibility---what they call the ``dynamic strength of ties''---obtained from the mentioned mapping.

It is worth noting that the speed of SI spreading---especially in the deterministic case where an infectious individual always infects a susceptible individual upon contact---is related to temporal path lengths, latency and reachability~\cite{holme_2005,pan_saramaki}, addressed elsewhere in this review. By placing constraints upon the timings between consecutive events defining a path, fastest temporal paths can also be representative of the pathways taken by the SIR spreading process~\cite{pan_saramaki}.

Finally, in addition to temporal inhomogeneities slowing down spreading in human communication networks, a similar effect has been observed for ants, however, compared to a different null model. In Ref.~\cite{blonder_dornhaus}, temporal networks were constructed based on 30-minute video recordings of ants in 6 colonies and tracking all contacts between individual ants. Compared to a kinetic null model, where ants were considered as gas particles that randomly collide and change directions, SI-like information flow was observed to be significantly slower for the empirical contact sequences at long time scales.

We also mention the modeling work by Kamp~\cite{kamp}, where the author proposes a framework to study disease spreading in temporal networks, which simulates the contagion process without explicitly generating interval graphs. In Kamp's setup, vertices come in to the system with a degree sampled from a degree distribution (which then changes due to the evolution of the graph); they live for a limited time and their degree changes due to both the birth and death of neighbors, and rewiring of edges. This framework allows for some approximate analytical treatments with generating functions and differential equation modeling but seems to require numerical simulations for a full characterization.

\subsection{Utilizing temporal structure for disease control}

The structure of contact patterns not only affects the spreading of disease, but this structure can also be exploited in controlling and preventing the spread. The most common preventive intervention is vaccination that lowers the probability of people catching the disease and spreading it further. Typically, one does not have to vaccinate the entire population to block the possibility of outbreaks. Already at some partial coverage $f$ of vaccinees in the population, disease cannot any longer propagate. This effect is called \emph{herd immunity}. Lowering the threshold of herd immunity is an important goal in public health. In static network theory, there are methods that utilize the network topology to identify important targets for vaccination. Perhaps the most well-known is the neighborhood vaccination protocol of Cohen \textit{et al.}~\cite{cohen_etal} that works by repeating the following steps: 
\begin{enumerate}
\item Take a random person in the population.
\item Ask the person to name a friend (or rather someone that person meets regularly in such a way that the disease in question might spread).
\item Vaccinate the friend.
\end{enumerate}
These steps are repeated until the desired fraction $f$ of the population is vaccinated. The benefits of this scheme are twofold. First, it utilizes only local information---a person is expected to know his or her own contacts, not any third person's contacts---which really is a prerequisite rather than just an advantage. Second, it samples people in proportion to their degree $k$. The importance of a person with degree $k$ with respect to disease spreading is, in classic network epidemiology, proportional to $k^2$---roughly speaking, the probability that the person gets the disease is proportional to $k$, and the expected number of other people the person infects is also proportional to $k$~\cite{anderson_may}. In a temporal network, it is important to remember that future contacts cannot be used to determine the right targets for vaccination, and that past contacts cannot benefit from current vaccination, unless the contacts are repeated. The basic assumption of classic network epidemiology---that the network is static---means that the past contacts will always also persist into the future, and this does not hold in general in temporal networks.

Lee \textit{et al.}~\cite{lee_etal} proposed an extension of the neighborhood vaccination scheme to temporal networks. They found that the strategy to ask about your most recent contact, or most frequent contact some time back in the past, improves the neighborhood vaccination in some real contact data sets (e.g.\ the prostitution sex network of Rocha \textit{et al.}~\cite{rocha_etal_2010}, the email data of Eckmann \textit{et al.}~\cite{eckmann_etal}, and the hospital proximity data of Liljeros \textit{et al.}~\cite{liljeros_etal_2007}). The response is different for different data sets, so for the hospital and prostitution data, the ``most recent''-version is the most efficient, whereas for the email data the most frequent is more effective. This, Lee \textit{et al.}\ argue, comes from the fact that the contacts along an edge in the hospital proximity and prostitution data are fairly limited in time---two people who enter into a period of frequent contacts in either one of these data sets will rather likely be in contact with other persons a bit later. For the email data, the driving force behind the efficiency of the ''most frequent''-protocol is that the contact frequency along an edge varies, but people in this dataset typically keep the relationship going throughout the data. From these simulations one can see that the temporal structure actually adds something that can be exploited to local vaccination programs. We believe other types of population-sampling protocols that are affected by network structure in static simulations could need to be extended to temporal networks.

\section{Future outlook}

In this review, we have illustrated how several systems can benefit from the temporal network approach, and discussed ways and methods of discovering and measuring network structure that lives in the time domain. Such structure becomes important for network studies especially in the context of some dynamics taking place on the network: if there are inhomogeneities and correlations in the contact sequences between vertices, these inhomogeneities may have dramatic effects on dynamics mediated through the contacts. The conceptual differences between modeling dynamical processes on static and temporal networks deserve some attention. For example, for simple contact processes such as disease spreading models, in the static network approach one typically integrates two components in the model: when the contacts along an edge take place, and the probability for the disease being transmitted during a contact between a susceptible and an infected vertex. The common assumption is that of contacts spread uniformly in time. In the temporal network approach, the first component---the timings of the contacts---is no longer a part of the spreading model, but rather an integral part of the contact structure, i.e.\ the network itself.  In general, for models of dynamical processes taking place on networks, moving from the static network framework to that of temporal networks is equivalent to removing the temporal component related to contact timings from the model---such a component is part of almost every model, although it may not be explicitly visible---and considering temporal contact structure instead.  But is the temporal network approach then nothing more than shifting a component of a dynamical model from the dynamical system to the underlying representation of interaction structure? What do we gain from such an approach, where the representation of the system in question necessarily becomes more complicated? First, as we have seen in this review, the temporal structure is crucial for studies of dynamical processes. Luckily, at least for processes involving humans, 
much of the data produced by today's technologies---be it mobile phone records~\cite{onnela_etal,karsai_etal} or wearable sensors~\cite{eagle_pentland,cattuto_etal}---comes in the form of contact sequences and are thus directly suitable for temporal network studies. Thus one can go straight from the raw data to simulations of spreading processes (or some other dynamics), and analyze the role of the temporal and topological structure by comparing the results to reference models~\cite{rocha_etal_2011}. Second, one may learn a lot about the system and its driving forces by studying its temporal patterns and structure, from inter-contact statistics to patterns involving multiple vertices such as motifs. Additionally, temporal and topological structures can be correlated~\cite{nordvik_liljeros}, and modeling this with an underlying static network structure is not straightforward.

The study of temporal networks, their characteristic features, and their dynamics is still a rather young field, and there are many open questions and unexplored directions. Below, we list some of these issues:

\emph{Generative models for temporal networks}. There are only very few models for temporal networks and their contact sequences, and one of the important open issues is clearly constructing and studying parametrized, generative models of temporal networks, e.g.\ of human contact sequences with their characteristic features observed in real-world data, such as skewed inter-contact time distributions, bursty dynamics, and circadian and weekly rhythms.

\emph{Measures for temporal network structure}. Although a large number of measures and characteristics for temporal graphs have been discussed in this review, they have mostly been generalizations of static network measures, and we feel that there is much room for improvements, e.g.\ in relation to simple measures of time-domain correlations of contact sequences. As an example, why does randomizing the order of contacts in a temporal network often, but not always, make spreading dynamics slower? It would be a little breakthrough if this question could be answered in terms of a simple, easily observable measure (akin to the statement ``high clustering coefficient slows down disease spreading'') \cite{szendroi_csanyi}. In a sense, recent studies on temporal networks are going in the opposite direction from the early work on complex networks during the millennium's first years~\cite{newman_2010}---beginning with the effects of the temporal structure on dynamical systems, whatever this structure may be, rather than first quantifying and characterizing the structural features of real systems, and only afterwards investigating the role of particular structural features on dynamics. Today, we know that temporal network structure can make a difference, but not exactly how or why.

\emph{Understanding the driving mechanisms}. A third, largely unexplored theme is why contacts between two vertices in a temporal network happen when they happen---there are skewed inter-contact time distributions along individual edges, but in general, why? There are many papers about why two vertices become connected by an edge~\cite{newman_2010} and there are papers explaining time series of when a vertex does something~\cite{barabasi}. However, presumably, what other vertices vertex $i$ is connected to could affect the times when $i$ does something, and vice versa. This question comes close to the goal of adaptive network studies~\cite{gross_blasius} that model the feedback from network structure (and how it affects dynamics on the network) to the success of the agents forming the network (and how they seek to change their position in it). If one could include when contacts happen along an edge into adaptive network models and thereby explain some observed temporal-topological correlations, this would be a breakthrough (no matter what the objective system is).

\emph{Inference problems}. Another set of more statistics-related challenges concerns inference problems. How can one construct a temporal network from various amounts of information about the states of vertices or edges~\cite{adar_adamic}? How can one infer spreading chains, if one has an incomplete temporal network? Many inference problems on static networks should be rather different on temporal graphs as many fundamental properties---like the transitivity of edges or Menger's theorem~\cite{berman}---are only valid under rather strong assumptions.

\emph{dynamical systems}. If a system benefit from being modeled as a temporal network is not only a question about the structure of the contacts, but also about the nature of the dynamical system acting over the contacts. Some dynamical systems might be more or less sensitive to temporal effects. In social information spreading, a person may, hypothetically, spread information only if he, or she, hears about it from two independent sources within a short period of time. Such a dynamical system should be sensitive to temporal effects like activity bursts and the order of events. 

\emph{Community, cluster or mesoscopic structure}. The recent years have seen tremendous efforts for discovering mesoscopic structure in static networks in the form of communities~\cite{fortunato}, loosely defined as groups of vertices more densely connect within than between each other. Most of the literature on community structure of static network focus on deriving a method for decomposing the network from some kind of conceptually simple principle. Few, if any, studies seeks to identify structures known \textit{a priori} to exist. The works incorporating a time dimension into the community detection (like Refs.~\cite{mucha_etal,rosvall_bergstrom,palla_etal, Ronhovde11}) operate on aggregated time-slices of the temporal network. One can imagine clustering algorithms based on more elaborate temporal structures, like time-respecting paths (a rare exception is Ref.~\cite{lin_etal}).

\emph{Visualization.} Visualization software is a great help for investigating static networks. Even though small graphs are impossible to embed in Euclidean space such that the Euclidean distance between vertices is proportional to the graph distance (that would be ideal for a direct correspondence between the real graph---the terrain---and the visualization---the map), still one can make visualizations that capture much of the network structure. Typically they organize the vertices such that there is a positive correlation between the graph and Euclidean distances~\cite{hachul_junger}. This is enough to see differences between networks of different degree--degree correlations~\cite{newman_2010} or to identify dense clusters by the eye~\cite{fortunato}. Stacking snapshots of a temporal network to a movie usually does a bad job to visualize temporal network structure, especially for sparse contact sequences. The time-line plots of e.g.\ Figs.~\ref{fig:reachability}, \ref{fig:tantipathananandh} and \ref{fig:randomization} resolve the vertices in one dimension, which makes it even harder to put vertices that are close in terms of latency (or some other distance-like metric for temporal networks) close to each other in Euclidean space. If one could find another clever way to layout a temporal network that captured reachability and latency that would be a very valuable contribution.

These open directions mentioned above mainly concern theoretical and methodological developments. However, the real acid test of temporal networks as a fruitful paradigm is its application to concrete, specific problems in population biology, cell biology, ecology, neuroscience, social and political sciences, economics, chemistry and so forth. So far, the temporal network framework has, with not so many exceptions, been investigated theoretically rather than used to explain the world around us. Yet most complex systems in the world are time-dependent, dynamical and in motion. As we believe that the temporal networks framework is really a tool for advancing science, we hope to see theoretically-minded researchers bringing it into collaborations with their applied colleagues.

\begin{acknowledgments}
The authors acknowledge financial support by the Swedish Research Council (PH), the WCU program through NRF Korea funded by MEST R31--2008--10029 (PH), EU's 7th Framework Program's FET-Open to ICTeCollective project no.\ 238597 (JS) and the Academy of Finland, the Finnish Center of Excellence program 2006--2011, project no.\ 129670 (JS). The authors thank (in alphabetic order) Albert-L\'aszlo Barab\'asi, Ginestra Bianconi, Ciro Cattuto, Aaron Clauset, Mikael Huss, Beom Jun Kim, Mario Konschake, Mirjam Kretzschmar, Vito Latora, Sune Lehmann, Jure Leskovec, Cecilia Mascolo, Esteban Moro, Zohar Nussinov, Etsuko Nonaka, John Tang, Robert Ulanowicz and Tao Zhou for comments.
\end{acknowledgments}


\begin{thebibliography}{100}

\bibitem{adar_adamic}
E.~Adar and L.~A. Adamic.
\newblock Tracking information epidemics in blogspace.
\newblock In {\em Proceedings of the 2005 IEEE/WIC/ACM International Conference
  on Web Intelligence}, pages 207--214, 2005.

\bibitem{alon}
U.~Alon.
\newblock Network motifs: Theory and experimental approaches.
\newblock {\em Nature Review Genetics}, 8:450--461, 2007.

\bibitem{anderson_may}
R.~A. Anderson and R.~A. May.
\newblock {\em Infectious diseases in human}.
\newblock Oxford University Press, Oxford UK, 1991.

\bibitem{paolo_etal}
P.~Bajardi, A.~Barrat, F.~Natale, L.~Savini, and V.~Colizza.
\newblock Dynamical patterns of cattle trade movements.
\newblock {\em PLoS ONE}, 6:e19869, 2011.

\bibitem{bajardi2011}
P.~Bajardi, A.~Barrat, F.~Natale, L.~Savini, and V.~Colizza.
\newblock Dynamical patterns of cattle trade movements.
\newblock {\em PLoS One}, 6:e19869, 2011.

\bibitem{bansal_etal}
S.~Bansal, J.~Read, B.~Pourbohloul, and L.~A. Meyers.
\newblock The dynamic nature of contact networks in infectious disease
  epidemiology.
\newblock {\em Journal of Biological Dynamics}, 4:478--489, 2010.

\bibitem{barabasi}
A.-L. Barab\'asi.
\newblock The origin of bursts and heavy tails in humans dynamics.
\newblock {\em Nature}, 435:207--212, 2005.

\bibitem{barrat_etal_2004}
A.~Barrat, M.~Barth\'elemy, R.~Pastor-Satorras, and A.~Vespignani.
\newblock The architecture of weighted complex networks.
\newblock {\em Proc. Natl. Acad. Sci. USA}, 101:3747, 2004.

\bibitem{barrat_etal_2008}
A.~Barrat, M.~Barth\'elemy, and A.~Vespignani.
\newblock {\em Dynamical processes on complex networks}.
\newblock Cambridge University Press, Cambridge UK, 2008.

\bibitem{barthelemy}
M.~Barth\'elemy.
\newblock Spatial networks.
\newblock {\em Physics Reports}, 499:1--101, 2011.

\bibitem{barthelemy_etal}
M.~Barth\'elemy, A.~Barrat, R.~Pastor-Satorras, and A.~Vespignani.
\newblock Velocity and hierarchical spread of epidemic outbreaks in scale-free
  networks.
\newblock {\em Physical Review Letters}, 92:178701, 2004.

\bibitem{bassett_etal}
D.~S. Bassett, N.~F. Wymbs, M.~A. Porter, P.~J. Mucha, J.~M. Carlson, and S.~T.
  Grafton.
\newblock Dynamic reconfiguration of human brain networks during learning.
\newblock {\em Proc. Natl. Acad. Sci. USA}, 108:7641--7646, 2011.

\bibitem{basu_etal}
P.~Basu, A.~Bar-Noy, R.~Ramanathan, and M.~P. Johnson.
\newblock Modeling and analysis of time-varying graphs.
\newblock e-print arXiv:1012.0260.

\bibitem{bearman_etal}
P.~Bearman, J.~Moody, and K.~Stovel.
\newblock Chains of affection: The structure of adolescent romantic and sexual
  networks.
\newblock {\em American Journal of Sociology}, 110:44--91, 2004.

\bibitem{berman}
K.~Berman.
\newblock Vulnerability of scheduled networks and a generalization of menger's
  theorem.
\newblock {\em Networks}, 28:125--134, 1996.

\bibitem{blonder_dornhaus}
B.~Blonder and A.~Dornhaus.
\newblock Time-ordered networks reveal limitations to information flow in ant
  colonies.
\newblock {\em PLoS ONE}, 6:e20298, 2011.

\bibitem{boguna_etal}
M.~Bogu{\~{n}}\'a, R.~Pastor-Satorras, and A.~Vespignani.
\newblock Epidemic spreading in complex networks with degree correlations.
\newblock In R.~Pastor-Satorras, M.~Rubi, and A.~Diaz-Guilera, editors, {\em
  Statistical mechanics of complex networks}, pages 127--147. Springer, Berlin,
  2003.

\bibitem{braha_bar_yam}
D.~Braha and Y.~Bar-Yam.
\newblock Time-dependent complex networks: dynamic centrality, dynamic motifs,
  and cycles of social interaction.
\newblock In T.~Gross and H.~Sayama, editors, {\em Adaptive networks: Theory,
  models and applications}, pages 39--50. Springer, Dordrecht, 2008.

\bibitem{buixuan_etal}
B.~{Bui Xuan}, A.~Ferreira, and A.~Jarry.
\newblock Computing shortest, fastest, and foremost journeys in dynamic
  network.
\newblock {\em International Journal of Foundations of Computer Science},
  14:267--285, 2002.

\bibitem{bullmore_sporns}
E.~Bullmore and O.~Sporns.
\newblock Complex brain networks: graph theoretical analysis of structural and
  functional systems.
\newblock {\em Nature Reviews Neuroscience}, 10:186, 2009.

\bibitem{candia_etal}
J.~Candia, M.~C. Gonz\'alez, P.~Wang, T.~Schoenharl, G.~Madey, and A.-L.
  Barab\'asi.
\newblock Uncovering individual and collective human dynamics from mobile phone
  records.
\newblock {\em Journal of Physics A}, 41:224015, 2008.

\bibitem{carley}
K.~M. Carley.
\newblock Dynamic network analysis.
\newblock In R.~Breiger, K.~M. Carley, and P.~Pattison, editors, {\em Dynamic
  Social Network Modeling and Analysis: Workshop Summary and Papers}, pages
  133--145. Committee on Human Factors, National Research Council, Washington
  DC, 2003.

\bibitem{casteigts_etal}
A.~Casteigts, P.~Flocchini, W.~Quattrociocchi, and N.~Santoro.
\newblock Time-varying graphs and dynamic networks.
\newblock In {\em Proceedings of the 10th International Conference on Adhoc
  Networks and Wireless (ADHOC-NOW)}, pages 346--359, 2011.

\bibitem{cattuto_etal}
C.~Cattuto, W.~{van den Broeck}, A.~Barrat, V.~Colizza, J.-F. Pinton, and
  A.~Vespignani.
\newblock Dynamics of person-to-person interactions from distributed {RFID}
  sensor networks.
\newblock {\em PLoS ONE}, 5:e11596, 2010.

\bibitem{chaintreau_etal}
A.~Chaintreau, A.~Mtibaa, L.~Massouli\'e, and C.~Diot.
\newblock Diameter of opportunistic mobile networks.
\newblock In {\em Proceedings of ACM Sigcomm CoNext}, 2007.

\bibitem{chechik_etal}
G.~Chechik, E.~Oh, O.~Rando, J.~Weissman, A.~Regev, and D.~Koller.
\newblock Activity motifs reveal principles of timing in transcriptional
  control of the yeast metabolic network.
\newblock {\em Nature Biotechnology}, 26:1251--1259, 2008.

\bibitem{cheng_etal}
E.~Cheng, J.~W. Grossman, and M.~J. Lipman.
\newblock Time-stamped graphs and their associated influence digraphs.
\newblock {\em Discrete Applied Mathematics}, 128:317--335, 2003.

\bibitem{clauset_eagle}
A.~Clauset and N.~Eagle.
\newblock Persistence and periodicity in a dynamic proximity network.
\newblock In {\em DIMACS Workshop on Computational Methods for Dynamic
  Interaction Networks}, Piscataway NJ, 2007. DIMACS.

\bibitem{cohen_etal}
R.~Cohen, S.~Havlin, and D.~Ben-Avraham.
\newblock Efficient immunization strategies for computer networks and
  populations.
\newblock {\em Phys. Rev. Lett.}, 91:247901, 2003.

\bibitem{cookehalsey}
K.~L. Cooke and E.~Halsey.
\newblock The shortest route through a network with time-dependent internodal
  transit times.
\newblock {\em Journal of Mathematical Analysis and Applications}, page 493,
  1966.

\bibitem{croft_etal}
D.~P. Croft, J.~Krause, and R.~James.
\newblock Social network in the guppy ({P}oecilia reticulata).
\newblock {\em Proc. R. Soc. B.}, 271:S516--S519, 2004.

\bibitem{daCostaSurvey}
L.~da~Fontoura~Costa, F.~A. Rodriguez, G.~Travieso, and P.~Villas~Boas.
\newblock Characterization of complex networks: {A} survey of measurements.
\newblock {\em Advances in Physics}, 56:167--242, 2007.

\bibitem{dagum_etal}
P.~Dagum, A.~Galper, and E.~Horvitz.
\newblock Dynamic network models for forecasting.
\newblock In {\em Proceedings of the eighth conference on Uncertainty in
  Artificial Intelligence}, pages 41--48, 1992.

\bibitem{deruiter_etal}
P.~C. {de Ruiter}, V.~Wolters, and J.~C. Moore, editors.
\newblock {\em Dynamic Food Webs: Multispecies Assemblages, Ecosystem
  Development and Environmental Change}.
\newblock Academic Press, London, 2005.

\bibitem{devicofallani_etal}
F.~{de Vico Fallani}, V.~Latora, L.~Astolfi, F.~Cincotti, D.~Mattia, M.~G.
  Marciani, S.~Salinari, A.~Colosimo, and F.~Babiloni.
\newblock Persistent patterns of interconnection in time-varying cortical
  networks estimated from high-resolution eeg recordings in humans during a
  simple motor act.
\newblock {\em J. Phys. A}, 41:224014, 2008.

\bibitem{dimitriadis_etal}
S.~I. Dimitriadis, N.~A. Laskaris, V.~Tsirka, M.~Vourkas, S.~Micheloyannis, and
  S.~Fotopoulos.
\newblock Tracking brain dynamics via time-dependent network analysis.
\newblock {\em Journal of Neuroscience Methods}, 193:145, 2010.

\bibitem{eagle_pentland}
N.~Eagle and A.~Pentland.
\newblock Reality mining: sensing complex social systems.
\newblock {\em Personal and Ubiquitous Computing}, 10:255--268, 2006.

\bibitem{easley_kleinberg}
D.~Easley and J.~Kleinberg.
\newblock {\em Networks, crowds, and markets: reasoning about a highly
  connected world}.
\newblock Cambridge University Press, Cambridge UK, 2010.

\bibitem{eckmann_etal}
J.-P. Eckmann, E.~Moses, and D.~Sergi.
\newblock Entropy of dialogues creates coherent structures in e-mail traffic.
\newblock {\em Proc. Natl. Acad. Sci. USA}, 101:14333--14337, 2004.

\bibitem{farrel}
A.~Farrel.
\newblock {\em The Internet and its protocols: A comparative approach}.
\newblock Elsevier, Amsterdam, 2004.

\bibitem{ferreira}
A.~Ferreira.
\newblock On models and algorithms for dynamic communication networks: The case
  for evolving graphs.
\newblock In {\em Proceedings of 4e rencontres francophones sur les Aspects
  Algorithmiques des T\'el\'ecommunications (ALGOTEL`2002)}, pages 155--161,
  M\`eze, 2002. INRIA Press.

\bibitem{fortunato}
S.~Fortunato.
\newblock Community detection in graphs.
\newblock {\em Physics Reports}, 486:75--174, 2010.

\bibitem{gautreau_etal}
A.~Gautreau, A.~Barrat, and M.~Barth\'elemy.
\newblock Microdynamics in stationary complex networks.
\newblock {\em Proc. Natl. Acad. Sci. USA}, 106:8847--8852, 2009.

\bibitem{ghosh}
S.~Ghosh.
\newblock {\em Distributed Systems: An Algorithmic Approach}.
\newblock Chapman \& Hall / CRC, Boca Raton FL, 2007.

\bibitem{Goh08}
K.-I. Goh and A.-L. Barab\'asi.
\newblock Burstiness and memory in complex systems.
\newblock {\em EPL}, 81:48002, 2008.

\bibitem{graciaPRE2011}
J.~F. Gracia, V.~M. Egu'iluz, and M.~San~Miguel.
\newblock Update rules and interevent time distributions: Slow ordering versus
  no ordering in the voter model.
\newblock {\em Physical Review E}, 84:015103, 2011.

\bibitem{grindrod_etal}
P.~Grindrod, M.~C. Parsons, D.~J. Higham, and E.~Estrada.
\newblock Communicability across evolving networks.
\newblock {\em Phys. Rev. E}, 81:046120, 2011.

\bibitem{gross_blasius}
T.~Gross and B.~Blasius.
\newblock Adaptive coevolutionary networks: A review.
\newblock {\em J. Roy. Soc. Interface}, 5:259--271, 2008.

\bibitem{gunturi_etal}
V.~Gunturi, S.~Shekhar, and A.~Bhattacharya.
\newblock Minimum spanning tree on spatio-temporal networks.
\newblock In {\em Proceedings of the 21th conference on database and expert
  systems application, part 2.}, pages 149--158, Heidelberg, 2010. Springer.

\bibitem{guo_etal}
F.~Guo, S.~Hanneke, W.~Fu, and E.~P. Xing.
\newblock Recovering temporally rewiring networks: A model-based approach.
\newblock In {\em International Conference of Machine Learning}, 2007.

\bibitem{hachul_junger}
S.~Hachul and M.~J\"unger.
\newblock An experimental comparison of fast algorithms for drawing general
  large graphs.
\newblock {\em Lecture Notes in Computer Science}, 3843:235--250, 2006.

\bibitem{han_etal}
J.-D.~J. Han, N.~Bertin, T.~Hao, D.~S. Goldberg, G.~F. Berriz, L.~V. Zhang,
  D.~Dupuy, A.~J.~M. Walhout, M.~E. Cusick, F.~P. Roth, and .~M. Vidali.
\newblock Evidence for dynamically organized modularity in the yeast
  proteinÐprotein interaction network.
\newblock {\em Nature}, 430:88--93, 2004.

\bibitem{hanneke_xing}
S.~Hanneke and E.~P. Xing.
\newblock Discrete temporal models of social networks. workshop on statistical
  network analysis.
\newblock In {\em Proceedings of the 23rd International Conference on Machine
  Learning (ICML-SNA)}, 2006.

\bibitem{harary_gupta}
F.~Harary and G.~Gupta.
\newblock Dynamic graph models.
\newblock {\em Mathematical and Computer Modelling}, 25:79--88, 1997.

\bibitem{harris}
T.~E. Harris.
\newblock {\em The Theory of Branching Processes}.
\newblock Springer, Berlin, 2002.

\bibitem{hethcote}
H.~W. Hethcote.
\newblock The mathematics of infectious diseases.
\newblock {\em SIAM Review}, 42:599, 2000.

\bibitem{hill_braha}
S.~A. Hill and D.~Braha.
\newblock Dynamic model of time-dependent complex networks.
\newblock {\em Phys. Rev. E}, 82:046105, 2010.

\bibitem{holme_2003}
P.~Holme.
\newblock Network dynamics of ongoing social relationships.
\newblock {\em Europhys. Lett.}, 64:427--433, 2003.

\bibitem{holme_2005}
P.~Holme.
\newblock Network reachability of real-world contact sequences.
\newblock {\em Phys. Rev. E}, 71:046119, 2005.

\bibitem{holme_etal}
P.~Holme, C.~E. Edling, and F.~Liljeros.
\newblock Structure and time-evolution of an {I}nternet dating community.
\newblock {\em Social Networks}, 26:155--174, 2004.

\bibitem{iribarren_moro}
J.~L. Iribarren and E.~Moro.
\newblock Impact of human activity patterns on the dynamics of information
  diffusion.
\newblock {\em Phys. Rev. Lett.}, 103:038702, 2009.

\bibitem{iribarren2011}
J.~L. Iribarren and E.~Moro.
\newblock Branching dynamics of viral information spreading.
\newblock e-print arXiv:1110.1884, 2011.

\bibitem{Isella_PLoS1}
L.~Isella, M.~Romano, A.~Barrat, C.~Cattuto, V.~Colizza, W.~Van~den Broeck,
  F.~Gesualdo, E.~Pandolfi, L.~Ravˆ, C.~Rizzo, and A.~E. Tozzi.
\newblock Close encounters in a pediatric ward: Measuring face-to-face
  proximity and mixing patterns with wearable sensors.
\newblock {\em PLoS ONE}, 6:e17144, 2011.

\bibitem{isella_etal}
L.~Isella, J.~Stehl\'e, A.~Barrat, C.~Cattuto, J.-F. Pinton, and W.~{Van den
  Broeck}.
\newblock WhatÕs in a crowd? analysis of face-to-face behavioral networks.
\newblock {\em Journal of Theoretical Biology}, 271:166--180, 2011.

\bibitem{jackson}
M.~O. Jackson.
\newblock {\em Social and economic networks}.
\newblock Princeton University Press, Princeton NJ, 2008.

\bibitem{java_etal}
A.~Java, X.~Song, T.~Finin, and B.~Tseng.
\newblock Why we twitter: Understanding microblogging usage and communities.
\newblock In {\em Proceedings of the 9th WebKDD and 1st SNA-KDD workshop on web
  mining and social network analysis}, 2007.

\bibitem{jo_etal}
H.-H. Jo, M.~Karsai, J.~Kert\'esz, and K.~Kaski.
\newblock Circadian pattern and burstiness in human communication activity.
\newblock e-print arXiv:1101.0377.

\bibitem{hangPlosOne2011}
H.-H. Jo, R.~K. Pan, and K.~Kaski.
\newblock Emergence of bursts and communities in evolving weighted networks.
\newblock {\em PLoS ONE}, 6:e22687, 2011.

\bibitem{johansen}
A.~Johansen.
\newblock Probing human response times.
\newblock {\em Physica A}, 330:286--291, 2004.

\bibitem{kamp}
C.~Kamp.
\newblock Untangling the interplay between epidemic spread and transmission
  network dynamics.
\newblock {\em PLoS Comp. Biol.}, 6:e1000984, 2010.

\bibitem{karsai_etal}
M.~Karsai, M.~Kivelä, R.~K. Pan, K.~Kaski, J.~Kert\'esz, A.~L. Barab\'asi, and
  J.~Saram\"aki.
\newblock Small but slow world: How network topology and burstiness slow down
  spreading.
\newblock {\em Phys. Rev. E}, 83:025102, 2011.

\bibitem{kauppi_etal}
J.-P. Kauppi, I.~J\"a\"askel\"ainen, M.~Sams, and J.~Tohka.
\newblock Inter-subject correlation of brain hemodynamic responses during
  watching a movie: localization in space and frequency.
\newblock {\em Journal of Neuroinformatics}, 4:5, 2009.

\bibitem{kempe_etal}
D.~Kempe, J.~Kleinberg, and A.~Kumar.
\newblock Connectivity and inference problems for temporal networks.
\newblock {\em Journal of Computer and System Sciences}, 64:820, 2002.

\bibitem{kenah_robins}
E.~Kenah and J.~M. Robins.
\newblock Second look at the spread of epidemics on networks.
\newblock {\em Physical Review E}, 76:036113, 2007.

\bibitem{kimmel_axelrod}
M.~Kimmel and D.~E. Axelrod.
\newblock {\em Branching Process in Biology}.
\newblock Springer, New York, 2002.

\bibitem{kleinberg}
J.~Kleinberg.
\newblock Bursty and hierarchical structure in streams.
\newblock {\em Data Mining and Knowledge Discovery}, 7:373--397, 2003.

\bibitem{kolar_etal}
M.~Kolar, L.~Song, A.~Ahmed, and E.~P. Xing.
\newblock Estimating time-varying networks.
\newblock {\em Annals of Applied Statistics}, 4:94--123, 2010.

\bibitem{komuro_white}
K.~Komurov and M.~White.
\newblock Revealing static and dynamic modular architecture of the eukaryotic
  protein interaction network.
\newblock {\em Molecular Systems Biology}, 3:110, 2007.

\bibitem{kossinets_etal}
G.~Kossinets, J.~Kleinberg, and D.~J. Watts.
\newblock The structure of information pathways in a social communication
  network.
\newblock In {\em Proc. 14th ACM SIGKKD Intl. Conf. on Knowledge Discovery and
  Data Mining}, pages 435--443, 2008.

\bibitem{kostakos}
V.~Kostakos.
\newblock Temporal graphs.
\newblock {\em Physica A}, 388:1007--1023, 2009.

\bibitem{Kovanen2011}
L.~Kovanen, M.~Karsai, K.~Kaski, J.~Kert\'esz, and J.~Saram\"aki.
\newblock Temporal motifs in time-dependent networks.
\newblock e-print arXiv:1107.5646.

\bibitem{kretzschmar_etal}
M.~Kretzschmar and M.~Morris.
\newblock Measures of concurrency in networks and the spread of infectious
  disease.
\newblock {\em Math. Biosci.}, 133:165--195, 1996.

\bibitem{kuhn_oshman}
F.~Kuhn and R.~Oshman.
\newblock Dynamic networks: Models and algorithms.
\newblock {\em ACM SIGACT News}, 42:82--96, 2011.

\bibitem{kumar_etal}
R.~Kumar, J.~Novak, P.~Raghavan, and A.~Tomkins.
\newblock On the bursty evolution of blogspace.
\newblock In {\em Proceedings of the 12th international conference on World
  Wide Web}, 2003.

\bibitem{Kumpula07}
J.~M. Kumpula, J.~P. Onnela, J.~Saram\"{a}ki, K.~Kaski, and J.~Kert\'{e}sz.
\newblock {Emergence of Communities in Weighted Networks}.
\newblock {\em Physical Review Letters}, 99:228701+, 2007.

\bibitem{kuwata_etal}
Y.~Kuwata, L.~Blackmore, M.~Wolf, N.~Fathpour, C.~Newman, and A.~Elfes.
\newblock Decomposition algorithm for global reachability analysis on a
  time-varying graph with an application to planetary exploration.
\newblock In {\em International Conference on Intelligent Robots and Systems},
  2009.

\bibitem{kwak_etal}
H.~Kwak, C.~Lee, H.~Park, and S.~Moon.
\newblock What is {T}witter, a social network or a news media?
\newblock In {\em Proceedings of the 19th International World Wide Web
  Conference}, 2010.

\bibitem{lahiri_berger_wolf_2007}
M.~Lahiri and T.~Y. Berger-Wolf.
\newblock Structure prediction in temporal networks using frequent subgraphs.
\newblock In {\em IEEE Symposium on Computational Intelligence and Data
  Mining}, pages 35--42, 2007.

\bibitem{lahiri_berger_wolf_2008}
M.~Lahiri and T.~Y. Berger-Wolf.
\newblock Mining periodic behavior in dynamic social networks.
\newblock In {\em Eighth IEEE International Conference on Data Mining}, 2008.

\bibitem{lamport}
L.~Lamport.
\newblock Time, clocks, and the ordering of events in a distributed system.
\newblock {\em Comm. ACM}, 21:558--565, 1978.

\bibitem{lebre}
S.~L\`ebre.
\newblock Inferring dynamic bayesian network with low order independencies.
\newblock {\em Statistical Applications in Genetics and Molecular Biology},
  8:9, 2009.

\bibitem{lebre_etal}
S.~L\`ebre, J.~Becq, F.~Devaux, M.~P.~H. Stumpf, and G.~Lelandais.
\newblock Statistical inference of the time-varying structure of
  gene-regulation networks.
\newblock {\em BMC Systems Biology}, 4:130, 2010.

\bibitem{lee_etal}
S.~Lee, L.~E.~C. Rocha, F.~Liljeros, and P.~Holme.
\newblock Exploiting temporal network structures of human interaction to
  effectively immunize populations.
\newblock e-print arXiv:1011.3928.

\bibitem{leland_wilson}
W.~E. Leland and D.~V. Wilson.
\newblock High time-resolution measurement and analysis of {LAN} traffic:
  Implications for lan interconnection.
\newblock In {\em Proceedings of InfoCom91}, pages 1360--1366, 1991.

\bibitem{leskovec_horvitz}
J.~Leskovec and E.~Horvitz.
\newblock Planetary-scale views on a large instant-messaging network.
\newblock In {\em Proceedings of the 17th International World Wide Web
  Conference}, pages 915--924, 2008.

\bibitem{liben_nowell_kleinberg}
D.~Liben-Nowell and J.~Kleinberg.
\newblock Tracing information flow on a global scale using {I}nternet
  chain-letter data.
\newblock {\em Proc. Natl. Acad. Sci. USA}, 105:4633--4638, 2008.

\bibitem{liljeros_edling_amaral}
F.~Liljeros, C.~E. Edling, and L.~A.~N. Amaral.
\newblock Sexual networks: Implications for the transmission of sexually
  transmitted infections.
\newblock {\em Microbes and Infections}, 5:189Ð196, 2003.

\bibitem{liljeros_etal_2001}
F.~Liljeros, C.~E. Edling, L.~A.~N. Amaral, H.~E. Stanley, and Y.~{\AA}berg.
\newblock The web of human sexual contacts.
\newblock {\em Nature}, 411:907--908, 2001.

\bibitem{liljeros_etal_2007}
F.~Liljeros, J.~Giesecke, and P.~Holme.
\newblock The contact network of inpatients in a regional health care system: a
  longitudinal case study.
\newblock {\em Mathematical Population Studies}, 14:269--284, 2007.

\bibitem{lin_etal}
Y.-R. Lin, Y.~Chi, S.~Zhu, H.~Sundaram, and B.~L. Tseng.
\newblock Facetnet: a framework for analyzing communities and their evolutions
  in dynamic networks.
\newblock In {\em Proceedings of the 17th international conference on World
  Wide Web}, pages 685--694, 2008.

\bibitem{lusseau_etal}
D.~Lusseau, K.~Schneider, O.~J. Boisseau, P.~Haase, E.~Slooten, and S.~M.
  Dawson.
\newblock The bottlenose dolphin community of {Doubtful Sound} features a large
  proportion of long-lasting associations.
\newblock {\em Behavioral Ecology and Sociobiology}, 54:396--405, 2003.

\bibitem{malmgren_etal_2009}
R.~D. Malmgren, D.~B. Stouffer, A.~S. L.~O. Campanharo, and L.~A.~N. Amaral.
\newblock On universality in human correspondence activity.
\newblock {\em Science}, 325:1696--1700, 2009.

\bibitem{malmgren_etal_2008}
R.~D. Malmgren, D.~B. Stouffer, A.~E. Motter, and L.~A.~N. Amaral.
\newblock A poissonian explanation for heavy tails in e-mail communication.
\newblock {\em Proc. Natl. Acad. Sci. USA}, 105:18153--18158, 2008.

\bibitem{mattern}
F.~Mattern.
\newblock Virtual time and global states of distributed systems.
\newblock In {\em Workshop on Parallel and Distributed Algorithms.}, 1989.

\bibitem{medo}
M.~Medo, G.~Cimini, and S.~Gualdi.
\newblock Temporal effects in the growth of networks.
\newblock e-print arXiv:1009.5560.

\bibitem{min_etal}
B.~Min, K.-I. Goh, and A.~Vazquez.
\newblock Spreading dynamics following bursty human activity patterns.
\newblock e-print arXiv:1006.2643.

\bibitem{miritello_etal}
G.~Miritello, E.~Moro, and R.~Lara.
\newblock The dynamical strength of social ties in information spreading.
\newblock {\em Phys. Rev. E}, 83:045102, 2011.

\bibitem{moody}
J.~Moody.
\newblock The importance of relationship timing for diffusion.
\newblock {\em Social Forces}, 81:25--56, 2002.

\bibitem{morris_kretzschmar}
M.~Morris and M.~Kretzschmar.
\newblock Concurrent partnerships and transmission dynamics in networks.
\newblock {\em Social Networks}, 17:299--318, 1995.

\bibitem{mucha_etal}
P.~J. Mucha, T.~Richardson, K.~Macon, M.~A. Porter, and J.-P. Onnela.
\newblock Community structure in time-dependent, multiscale, and multiplex
  networks.
\newblock {\em Science}, 328:876--878, 2010.

\bibitem{newman_2002}
M.~E.~J. Newman.
\newblock Spread of epidemic disease on networks.
\newblock {\em Phys. Rev. E}, 66:16128, 2002.

\bibitem{newman_2010}
M.~E.~J. Newman.
\newblock {\em Networks: An introduction}.
\newblock Oxford University Press, Oxford UK, 2010.

\bibitem{nicosia_etal}
V.~Nicosia, J.~Tang, M.~Musolesi, G.~Russo, C.~Mascolo, and V.~Latora.
\newblock Components in time-varying graphs.
\newblock e-print arXiv:1106.2134.

\bibitem{nordvik_liljeros}
M.~K. Nordvik and F.~Liljeros.
\newblock Number of sexual encounters involving intercourse and the
  transmission of sexually transmitted infections.
\newblock {\em Sexually Transmitted Diseases}, 33:342--349, 2006.

\bibitem{oliveira_barabasi}
J.~G. Oliveira and A.-L. Barab\'asi.
\newblock Human dynamics: {D}arwin and {E}instein correspondence patterns.
\newblock {\em Nature}, 437:1251, 2005.

\bibitem{onnela_etal}
J.-P. Onnela, J.~Saram\"aki, J.~Hyv\"onen, G.~Szab\'o, D.~Lazer, K.~Kaski,
  J.~Kert\'esz, and A.-L. Barab\'asi.
\newblock Structure and tie strengths in mobile communication networks.
\newblock {\em Proc. Natl. Acad. Sci. USA}, 104:7332, 2007.

\bibitem{pahlwostl}
C.~Pahl-Wostl.
\newblock {\em The dynamic nature of ecosystems: chaos and order entwined}.
\newblock Wiley, Chichester UK, 1995.

\bibitem{palla_etal}
G.~Palla, A.-L. Barab\'asi, and T.~Vicsek.
\newblock Quantifying social group evolution.
\newblock {\em Nature}, 446:664--667, 2007.

\bibitem{palsson}
B.~{\O}. Palsson.
\newblock {\em Systems Biology: Properties of Reconstructed Networks}.
\newblock Cambridge University Press, Cambridge UK, 2006.

\bibitem{pan_saramaki}
R.~K. Pan and J.~Saram\"aki.
\newblock Path lengths, correlations, and centrality in temporal networks.
\newblock {\em Phys. Rev. E}, 84:016105, 2011.

\bibitem{panisson_etal}
A.~Panisson, A.~Barrat, C.~Cattuto, W.~{Van den Broeck}, G.~Ruffo, and
  R.~Schifanella.
\newblock On the dynamics of human proximity for data diffusion in ad-hoc
  networks.
\newblock to appear in Ad Hoc Networks, 2011.

\bibitem{park_etal}
Y.~Park, C.~Moore, and J.~S. Bader.
\newblock Dynamic networks from hierarchical {B}ayesian graph clustering.
\newblock {\em PLoS ONE}, 5:e8118, 2010.

\bibitem{pascual_dunne}
M.~Pascual and J.~Dunne.
\newblock {\em Ecological Networks: Linking Structure to Dynamics in Food
  Webs}.
\newblock Oxford University Press, Oxford UK, 2006.

\bibitem{pastor_santorras_vespignani_2001}
R.~Pastor-Satorras and A.~Vespignani.
\newblock Epidemic spreading in scale-free networks.
\newblock {\em Phys. Rev. Lett.}, 86:3200--3203, 2001.

\bibitem{pastor_santorras_vespignani_2004}
R.~Pastor-Satorras and A.~Vespignani.
\newblock {\em Evolution and Structure of the Internet: A Statistical Physics
  Approach}.
\newblock Cambridge University Press, Cambridge UK, 2004.

\bibitem{przytycka_etal}
T.~M. Przytycka and M.~S. D.~K. Slonim.
\newblock Toward the dynamic interactome: It's about time.
\newblock {\em Briefings in Bioinformatics}, 11:15--29, 2010.

\bibitem{rao_hero_etal}
A.~Rao, A.~O. {Hero III}, D.~J. States, and J.~D. Engel.
\newblock Inferring time-varying network topologies from gene expression data.
\newblock {\em EURASIP Journal on Bioinformatics and Systems Biology},
  2007:51947, 2007.

\bibitem{riolo_etal}
C.~S. Riolo, J.~S. Koopman, and J.~S. Chick.
\newblock Methods and measures for the description of epidemiological contact
  networks.
\newblock {\em Journal of Urban Health}, 78:446--457, 2001.

\bibitem{robins_etal}
G.~Robins, P.~Pattison, Y.~Kalish, and D.~Lusher.
\newblock An introduction to exponential random graph models for social
  networks.
\newblock {\em Social Networks}, 29:173--191, 2006.

\bibitem{rocha_etal_2010}
L.~E.~C. Rocha, F.~Liljeros, and P.~Holme.
\newblock Information dynamics shape the sexual networks of internet-mediated
  prostitution.
\newblock {\em Proc. Natl. Acad. Sci. USA}, 107:5706--5711, 2010.

\bibitem{rocha_etal_2011}
L.~E.~C. Rocha, F.~Liljeros, and P.~Holme.
\newblock Simulated epidemics in an empirical spatiotemporal network of 50,185
  sexual contacts.
\newblock {\em PloS Comp. Biol.}, 7:e1001109, 2011.

\bibitem{Ronhovde11}
P.~Ronhovde, S.~Chakrabarty, D.~Hu, M.~Sahu, K.~F. Kelton, N.~A. Mauro, K.~K.
  Sahu, and Z.~Nussinov.
\newblock Detecting hidden spatial and spatio-temporal structures in glasses
  and complex physical systems by multiresolution network clustering.
\newblock 2011.
\newblock eprint arXiv:1102.1519.

\bibitem{rosvall_bergstrom}
M.~Rosvall and C.~T. Bergstrom.
\newblock Mapping change in large networks.
\newblock {\em PLoS ONE}, 5:e8694, 2010.

\bibitem{santoro_etal}
N.~Santoro, W.~Quattrociocchi, P.~Flocchini, A.~Casteigts, and F.~Amblard.
\newblock Time-varying graphs and social network analysis: temporal indicators
  and metric.
\newblock In {\em Proceedings of the 3rd AISB Social Networks and Multiagent
  Systems Symposium (SNAMAS)}, pages 32--38, 2011.

\bibitem{snijders_koskinen_etal}
T.~A.~B. Snijders, J.~Koskinen, and M.~Schweinberger.
\newblock Maximum likelihood estimation for social network dynamics.
\newblock {\em The Annals of Applied Statistics}, 4:567--588, 2010.

\bibitem{snijders_vandebunt_etal}
T.~A.~B. Snijders, G.~G. {van de Bunt}, and C.~E.~G. Steglich.
\newblock Introduction to stochastic actor-based models for network dynamics.
\newblock {\em Social Networks}, 32:44--60, 2010.

\bibitem{sole_bascompte}
R.~V. Sol\'e and J.~Bascompte.
\newblock {\em Self-Organization in Complex Ecosystems}.
\newblock Princeton University Press, Princeton NJ, 2006.

\bibitem{sporns_etal}
O.~Sporns, D.~R. Chialvo, M.~Kaiser, and C.~C. Hilgetag.
\newblock Organization, development and function of complex brain networks.
\newblock {\em Trends in Cognitive Sciences}, 8:418--425, 2004.

\bibitem{stehle_etal}
J.~Stehl\'e, A.~Barrat, and G.~Bianconi.
\newblock Dynamical and bursty interactions in social networks.
\newblock {\em Phys. Rev. E}, 81:035101, 2010.

\bibitem{Stehle:2011nx}
J.~Stehl\'{e}, N.~Voirin, A.~Barrat, C.~Cattuto, V.~Colizza, L.~Isella,
  C.~Regis, J.-F. Pinton, N.~Khanafer, W.~Van~den Broeck, and P.~Vanhems.
\newblock Simulation of an {SEIR} infectious disease model on the dynamic
  contact network of conference attendees.
\newblock {\em BMC Medicine}, 9(87), 2011.

\bibitem{stehle_etal_2011}
J.~Stehl\'e, N.~Voirin, A.~Barrat, C.~Cattuto, V.~Colizza, L.~Isella, C.~Regis,
  J.-F. Pinton, N.~Khanafer, W.~{Van den Broeck}, and P.~Vanhems.
\newblock Simulation of an seir infectious disease model on the dynamic contact
  network of conference attendees.
\newblock {\em BMC Medicine}, 9:87, 2011.

\bibitem{Stehle2011PLos1}
J.~Stehl\'{e}, N.~Voirin, A.~Barrat, C.~Cattuto, L.~Isella, J.-F. Pinton,
  M.~Quaggiotto, W.~Van~den Broeck, C.~RŽgis, B.~Lina, and P.~Vanhems.
\newblock High-resolution measurements of face-to-face contact patterns in a
  primary school.
\newblock {\em PLoS ONE}, 6:e23176, 2011.

\bibitem{sundaresan_etal}
S.~R. Sundaresan, I.~R. Fischhoff, J.~Dushoff, and D.~I. Rubenstein.
\newblock Network metrics reveal differences in social organization between two
  fission-fusion species, {G}revy's zebra and onager.
\newblock {\em Oecologia}, 151:140--149, 2006.

\bibitem{szendroi_csanyi}
B.~Szendroi and G.~Csanyi.
\newblock Polynomial epidemics and clustering in contact networks.
\newblock {\em Proc. Roy. Soc. B}, 271:S364--S366, 2004.

\bibitem{takaguchi_etal}
T.~Takaguchi, M.~Nakamura, N.~Sato, K.~Yano, and N.~Masuda.
\newblock Predictability of conversation patterns.
\newblock {\em Phys. Rev. X}, 1:011008, 2011.

\bibitem{tang_etal_2011}
J.~Tang, C.~Mascolo, M.~Musolesi, and V.~Latora.
\newblock Exploring temporal complex network metrics in mobile malware
  containment.
\newblock In {\em 12th IEEE International Symposium on a World of Wireless,
  Mobile and Multimedia Networks (WOWMOM `11)}, 2011.

\bibitem{tang_etal_2009}
J.~Tang, M.~Musolesi, C.~Mascolo, and V.~Latora.
\newblock Temporal distance metrics for social network analysis.
\newblock In {\em Proceedings of the 2nd ACM SIGCOMM Workshop on Online Social
  Networks}, page~3, 2009.

\bibitem{tang_etal_2010b}
J.~Tang, M.~Musolesi, C.~Mascolo, V.~Latora, and V.~Nicosia.
\newblock Analysing information flows and key mediators through temporal
  centrality metrics.
\newblock In {\em Proceeding of the 3rd ACM EuroSys Workshop on Social Networks
  Systems (SNS'10)}, page~3, 2010.

\bibitem{tang_etal_2010}
J.~Tang, S.~Scellato, M.~Musolesi, C.~Mascolo, and V.~Latora.
\newblock Small-world behavior in time-varying graphs.
\newblock {\em Phys. Rev. E}, 81:055101, 2010.

\bibitem{tantipathananandh_etal}
C.~Tantipathananandh, T.~Y. Berger-Wolf, and D.~Kempe.
\newblock A framework for community identification in dynamical social
  networks.
\newblock In {\em Proceedings of the 13th ACM SIGKDD International Conference
  on Knowledge Discovery and Data Mining}, pages 717--726, 2007.

\bibitem{taylor_etal}
I.~W. Taylor, R.~Linding, D.~Warde-Farley, Y.~Liu, C.~Pesquita, D.~Faria,
  S.~Bullamd, T.~Pawson, Q.~Morris, and J.~L. Wrana.
\newblock Dynamic modularity in protein interaction networks predicts breast
  cancer outcomes.
\newblock {\em Nature Biotech.}, 27:199--204, 2009.

\bibitem{timo_etal}
R.~Timo, K.~Blackmore, and L.~Hanlen.
\newblock On entropy measures for dynamic network topologies: {L}imits to
  {MANET}.
\newblock In {\em Proceedings of the sixth Australian Communications Theory
  Workshop}, pages 95--101, 2005.

\bibitem{turova}
T.~S. Turova.
\newblock Dynamical random graphs with memory.
\newblock {\em Phys. Rev. E}, 65:066102, 2002.

\bibitem{ueno_masuda}
T.~Ueno and N.~Masuda.
\newblock Controlling nosocomial infection based on structure of hospital
  social networks.
\newblock {\em J. Theor. Biol.}, 254:655--666, 2008.

\bibitem{ulanovicz}
R.~E. Ulanowicz.
\newblock Quantitative methods for ecological network analysis.
\newblock {\em Comp. Biol. Chem.}, 28:321--339, 2004.

\bibitem{valencia_etal}
M.~Valencia, J.~Martinerie, S.~Dupont, and M.~Chavez.
\newblock Dynamic small-world behavior in functional brain networks unveiled by
  an event-related networks approach.
\newblock {\em Phys. Rev. E}, 77:050905, 2008.

\bibitem{vazquez_etal}
A.~Vazquez, B.~R\'acz, A.~Luk\'acs, and A.-L. Barab\'asi.
\newblock Impact of non-poissonian activity patterns on spreading processes.
\newblock {\em Phys. Rev. Lett.}, 98:158702, 2007.

\bibitem{vernon_keeling}
M.~C. Vernon and M.~J. Keeling.
\newblock Representing the {UK's} cattle herd as static and dynamic networks.
\newblock {\em Proc. R. Soc. B}, 276:469--476, 2009.

\bibitem{volz_meyers}
E.~Volz and L.~A. Meyers.
\newblock Susceptible-infected-recovered epidemics in dynamic contact networks.
\newblock {\em Proc. Roy. Soc. B}, 274:2925--2934, 2007.

\bibitem{wasserman_faust}
S.~Wasserman and K.~Faust.
\newblock {\em Social Network Analysis: Methods and Applications}.
\newblock Cambridge University Press, Cambridge UK, 1994.

\bibitem{watts_strogatz}
D.~J. Watts and S.~H. Strogatz.
\newblock Collective dynamics of `small-world' networks.
\newblock {\em Nature}, 393:409--410, 1998.

\bibitem{wu_etal}
Y.~Wu, C.~Zhou, J.~Xiao, J.~Kurths, and H.~J. Schellnhuber.
\newblock Evidence for a bimodal distribution in human communication.
\newblock {\em Proc. Natl. Acad. Sci. USA}, 107:18803--18808, 2010.

\bibitem{zimo}
Z.~Yang, A.-X. Cui, and T.~Zhou.
\newblock Impact of heterogeneous human activities on epidemic spreading.
\newblock {\em Physica A}, 390:4543--4548, 2011.

\bibitem{yasseri11}
T.~Yasseri, R.~Sumi, and J.~Kert\'{e}sz.
\newblock Circadian patterns of {W}ikipedia editorial activity: {A} demographic
  analysis.
\newblock eprint arXiv:1109.1746, 2011.

\bibitem{yoshida_imoto_etal}
R.~Yoshida, S.~Imoto, and T.~Higuchi.
\newblock Estimating time-dependent gene networks from time series microarray
  data by dynamic linear models with markov switching.
\newblock In {\em CSB '05: Proceedings of the 2005 IEEE Computational Systems
  Bioinformatics Conference}, 289-298, 2005. IEEE Computer Society.

\bibitem{yoshida_etal}
T.~Yoshida, L.~E. Jones, S.~P. Ellner, G.~F. Fussmann, and N.~G. {Hairston Jr}.
\newblock Rapid evolution drives ecological dynamics in a predator-prey system.
\newblock {\em Nature}, 424:303--306, 2003.

\bibitem{zhao_etal_2011}
K.~Zhao, J.~Stehl\'e, G.~Bianconi, and A.~Barrat.
\newblock Social network dynamics of face-to-face interactions.
\newblock {\em Phys. Rev. E}, 83:056109, 2011.

\bibitem{zhao_etal}
Q.~Zhao, Y.~Tian, Q.~He, N.~Oliver, R.~Jin, and W.-C. Lee.
\newblock Communication motifs: A tool to characterize social communications.
\newblock In {\em Proceedings of the 19th ACM international conference on
  Information and knowledge management}, page 1645, 2010.

\bibitem{dan_etal}
Z.-D. Zhao, H.~Xia, M.-S. Shang, and T.~Zhou.
\newblock Empirical analysis on the human dynamics of a large-scale short
  message communication system.
\newblock {\em Chinese Physics Letters}, 28:068901, 2011.

\bibitem{zhou_etal}
T.~Zhou, Z.-D. Zhao, Z.~Yang, and C.~Zhou.
\newblock RelativeÊclockÊverifiesÊendogenousÊburstsÊofÊhumanÊdynamics.
\newblock e-print arXiv:1106.5562.

\end{thebibliography}
\end{document}